\documentclass{ar-1col}
 \usepackage[utf8]{inputenc}
 \usepackage[T1]{fontenc}
\usepackage[numbers]{natbib}
\usepackage{url}
\usepackage{hyperref}
\usepackage{subcaption}
\usepackage[table]{xcolor}

\jname{Annu. Rev. Nucl. Part. Sci.: Submitted}
\jvol{71}
\jyear{2020}
\doi{10.1146/annurev-nucl-102819-100428}

\begin{document}

\markboth{Henn}{What can we learn about QCD and collider physics from $\mathcal{N}=4$ super Yang-Mills?}

\title{What can we learn about QCD and collider physics from $\mathcal{N}=4$ super Yang-Mills?}

\author{Johannes M. Henn 
\affil{Max Planck Institute for Physics, Munich, Germany, 80805; email: henn@mpp.mpg.de}
}

\begin{abstract}
Tremendous ongoing theory efforts are dedicated to developing new methods for QCD calculations.
Qualitative rather than incremental advances are needed to fully exploit data still to be collected at the LHC.
The maximally supersymmetric Yang-Mills theory (${\mathcal N}=4$ sYM) shares with QCD the gluon sector, 
which contains the most complicated Feynman graphs, but at the same time has many special properties,
and is believed to be solvable exactly. It is natural to ask what we can learn from advances in ${\mathcal N}=4$ sYM for 
addressing difficult problems in QCD.
With this in mind, we review here several remarkable developments and highlights of recent results in ${\mathcal N}=4$ sYM. 
This includes all-order results for certain scattering amplitudes, novel symmetries, 
surprising geometrical structures of loop integrands, novel tools for the calculation of Feynman integrals, 
and bootstrap methods. While several insights and tools have already been carried over to QCD 
and have contributed to state-of-the-art calculations for LHC physics, we argue that there is a host of further fascinating ideas waiting to be explored.
\end{abstract}

\maketitle

\tableofcontents

\section{INTRODUCTION}

\subsection{Motivation}

Four-dimensional Yang-Mills theory, which is at the core of quantum chromodynamics (QCD), remains complicated, despite having been studied for more than half a century. Only with great efforts can theoretical predictions be made that keep up with the accuracy of experimental data collected at the LHC.
Compare this with developments in a close cousin of QCD, the maximally supersymmetric Yang-Mills theory, $\mathcal{N}=4$ super Yang-Mills (sYM).
Many fantastic advances made over the last two decades make many researchers think that, at least in the planar limit, the theory may be solved exactly!
The two theories share the Yang-Mills sector, so that tree-level gluon amplitudes are identical in both theories. 
Although the gluon amplitudes differ at loop level due to the matter content, the gluon diagrams, which are the most complicated ones, are the same.

What can we learn from progress in $\mathcal{N}=4$ sYM for QCD calculations?
In this review we wish to share the excitement about surprising and remarkable results, 
and to convey the conceptual and technological advances that led to them.
Moreover, we wish to point out where this research has already led to a transfer of knowledge to QCD. 
We also hope that by outlining recent developments, this review will help to promote positive exchange among the research communities.

We intend this review to be accessible (and hopefully enjoyable!) to non-expert readers from various fields of science,
including researchers and students from fields such as experimental physics or phenomenology,
or QCD practitioners, who are curious to know more about this subject.
Some readers may wish to get an overview of this research area to see if there are interconnections to other work.
Others may have heard buzzwords such as `transcendentality', `symbol', `amplituhedron', `bootstrap', and so on. 
In the following pages, we aim to explain those terms in a non-technical way.

Some readers might ask themselves whether studies in $\mathcal{N}=4$ sYM, however rewarding they may be in their own right, 
are not somewhat esoteric, in the sense that they seem far removed from the gritty calculations of `real' QCD. 
In some cases it can be beneficial to view QCD as a perturbation around $\mathcal{N}=4$ sYM, but this has limited scope.
Our viewpoint is rather that we can learn about new concepts in quantum field theory that would be hard to discover in a more complicated Yang-Mills theory. 
A particularly interesting topic is understanding physically motivated singular limits, such as the
important high-energy or collinear limits, where one often finds universal behavior.
New insights into $\mathcal{N}=4$ sYM have already led to novel tools for QCD, 
and are being used for cutting edge calculations relevant to LHC physics.
Beyond this, there is a host of further ideas and concepts available in the `$\mathcal{N}=4$ sYM world',
that have been considered only very recently, and whose potential for generalization to other theories is yet to be explored.
M.~Shifman \cite{Shifman:2008zr} describes this philosophy very clearly as follows:
{\it{
``Although the ultimate goal [...] is calculating QCD amplitudes, the concept design of various ideas and methods is carried out in supersymmetric theories, which provide an excellent testing ground. Looking at super-Yang-Mills offers a lot of insight into how one can deal with the problems in QCD.''
}
}

With this in mind here are a few concrete questions we think are important:
\vspace{-0.0cm}
\begin{itemize}
\item Can we develop methods to systematically compute Feynman integrals in QCD?
\item Can we compute physical quantities without explicitly evaluating Feynman diagrams?
\item How can our calculations benefit from knowledge of physical properties of the underlying QFT, 
such as unitarity, space-time symmetries, and conformal invariance?
\item Can we compute finite physical quantities in a way that avoids infrared singularities?
\item Which properties of QCD scattering amplitudes are governed by Wilson loops?
\end{itemize}

\subsection{Special properties of ${\mathcal N}=4$ super Yang-Mills}

The maximally supersymmetric Yang-Mills theory is often considered an idealized toy model for a possibly solvable four-dimensional Yang-Mills theory.
Its Lagrangian can be written most compactly in ten-dimensional notation,
\begin{equation}
\mathcal{L} = {\rm tr}\left( -\frac{1}{2} F_{MN} F^{MN} + i   \overline{\Psi} \Gamma^{N} \mathcal{D}\Psi \right) \,,
\end{equation}
where $\mathcal{D}$ is a covariant derivative, and $F^{MN}$ is the ten-dimensional field strength. 
The four-dimensional Lagrangian is then obtained by dimensional reduction, i.e. assuming that the fields only depend on the first four space-time dimensions.
Moreover, one writes the ten-dimensional gauge field as $A^{M} = (A^{\mu} , \phi^{i})$, where $A^{\mu}$ is the usual four-dimensional gauge field, and $\phi$ are six real scalars.
In this way one obtains a four-dimensional Lagrangian
that contains gluons coupled canonically to the scalars and to four complex fermions.
Furthermore, there are Yukawa and quartic scalar interaction vertices, in such a way as to preserve $\mathcal{N}=4$ supersymmetry, see e.g. \cite{Sohnius:1985qm}.
All fields are in the adjoint representation of $SU(N)$.
It is important to note that many of the methods discussed in this review do not rely on the explicit Lagrangian.

The theory has many special properties in addition to being supersymmetric. 
The combination of scale and Poincar\'e invariance of the Lagrangian lead to invariance under a larger continuous group, the conformal group, which includes translations, rotations and boosts, scale transformations, and coordinate inversions. In four dimensions, in which the Lorentz group is isomorphic to SO(3,1), the conformal group is isomorphic to SO(4,2). This leads to powerful constraints.
It is striking that this symmetry is not broken by quantum corrections, 
as the theory has a vanishing beta function to all orders in perturbation theory \cite{Mandelstam:1982cb,Howe:1983wj}.
\begin{marginnote}[]
\entry{Conformal group}{Extension of the Poincar\'e group to all transformations that preserve angles (in Euclidean space).}
\end{marginnote}

Moreover, the AdS/CFT correspondence conjectures a duality between $d$-dimensional conformal field theory and string theory in $(d+1)$-dimensional anti de-Sitter space \cite{Maldacena:1997re,Gubser:1998bc,Witten:1998qj}.
The duality comes with a `dictionary' that relates observables in the different theories. One of the best studied cases of the duality is $\mathcal{N}=4$ sYM on the one hand, and string theory on AdS${}_{5}$ space on the other hand.
The nature of this duality, which relates field theory at strong coupling to string theory at weak coupling, implies that the perturbative series must have special properties.
Remarkably simple structures have indeed been observed, e.g. for anomalous dimensions and especially in recent years, for scattering amplitudes.
Indeed, it appears that $\mathcal{N}=4$ sYM `wants' to teach us about nice mathematical structures, 
and all we need to do is to investigate interesting quantities in the theory and study their properties.

\subsection{Scope of the review and related pedagogical reviews}

The focus of this review is on developments related to scattering amplitudes. 
The reason for this is twofold. On the one hand, scattering amplitudes are obviously of interest for collider physics,
which is timely in view of the third run of the LHC. On the other hand, this has been and continues to be a particularly active
area in $\mathcal{N}=4$ sYM, and we think that there is considerable potential in applying insights from
that theory to QCD calculations. 

\begin{marginnote}[]
\entry{Scattering amplitudes}{Key ingredients of cross sections, analogous to probability amplitudes in quantum mechanics.}
\end{marginnote}

Given the wealth of results accumulated over many years, it is very difficult to make a selection.
It would have been possible to write a review four times this length, covering important topics in more
detail, and giving full justice to the many developments discussed for example at the yearly Amplitudes conferences.
A guiding principle was primarily to present developments that have potential for application in
more general settings, or are surprising, such as all-orders results for certain quantities in an interacting 
four-dimensional gauge theory.   

References to the original literature are given as much as possible to help readers learn more. 
We also point out a number of related resources, after the list of references. These include several review articles on scattering amplitudes, 
as well as on closely related research areas, such as integrability in planar ${\mathcal N}=4$ sYM, and conformal methods.

\subsection{Outline}

Sections 2 and 3 focus on selected highlights of exact results, with the intention of giving the reader a taste
of what may be possible in a four-dimensional gauge theory.
We then focus on perturbation theory, where one hopes to see the closest similarities with QCD.
Many explicit results for amplitudes are in some sense the tip of the iceberg, 
and in part are made possible and supported by a large body of work on loop integrands, which is the topic of section 4.
In section 5 we then discuss new techniques for Feynman integrals and for the special functions arising in them.
In section 6 we consider prospects for computing infrared finite Feynman integrals and observables.
The `bootstrap' method for higher loop amplitudes in section 7 brings together many of these ideas.
The conclusion follows in section 8.
At different places in the review there are wide blue boxes highlighting fascinating related subjects that are not discussed in the main text but may be of interest for further reading.

\section{THE CUSP ANOMALOUS DIMENSION}
\label{section-anomalousdimensions}

In quantum field theory, anomalous dimensions are important quantities. 
In QCD, the scaling evolution of parton distributions is governed by the anomalous dimension of composite operators.
In a conformally invariant field theory, scaling dimensions are fundamental because they govern the short-distance properties of correlation functions via the operator product expansion \cite{DiFrancesco:1997nk}.

\begin{marginnote}[]
\entry{{Cusp anomalous dimension}}{Determines the leading soft and collinear divergences of scattering amplitudes.}
\end{marginnote}

\begin{figure}[t]
\includegraphics[width=1in]{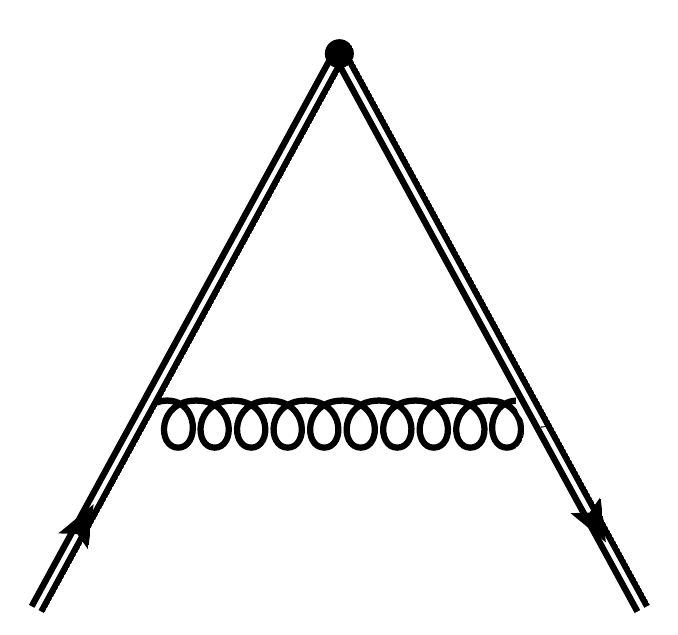}
\caption{One-loop Feynman diagram contributing to the vacuum expectation value of a Wilson loop formed by two segments.
The leading short-distance divergence defines the cusp anomalous dimension.
}
\label{figcusp1}
\end{figure}

The cusp anomalous dimension is of particular interest.
It is defined from the vacuum expectation value of certain Wilson loops,
\begin{equation}
\langle W_{C} \rangle = \langle {\rm tr} \, {\mathbf P} \exp \left( i g \oint_C A_{\mu} dx^{\mu} \right)\rangle \,, 
\end{equation}
where the integral is along a closed contour $C$, and ${\mathbf P}$ stands for path ordering of the $SU(N)$-valued field $A_{\mu}$ along the contour.
When the contour $C$ is not smooth, but contains a cusp as in Figure~\ref{figcusp1}, then there are short-distance divergences, which are controlled by the cusp anomalous dimension \cite{Polyakov:1980ca}. One can equivalently interpret the anomalous dimension as describing divergences due to soft gluon exchanges between two particles whose classical trajectories are given by the two segments of the contour, or due to virtual gluons emitted and absorbed by a particle that is hard-scattered at the point of the cusp. 

We discuss here the case where the cusp is formed by two null, or light-like, segments \cite{Korchemskaya:1992je}, 
and we denote the associated anomalous dimension by $\Gamma_{\rm cusp}$.
It depends on the Yang-Mills coupling $g_{YM}$, and on the rank $N$ of the gauge group $SU(N)$.

Its importance comes from the fact that it appears in many quantities. 
For example, it controls the large spin behavior of twist two operators \cite{Korchemskaya:1992je}, 
and it governs soft and collinear divergences of form factors and scattering amplitudes \cite{Collins:1989gx,Korchemsky:1993uz}.
It also plays a prominent role in the high-energy (Regge) limit \cite{Korchemskaya:1996je}, and more generally often appears in special singular limits.

In the following sections we will frequently discuss $\mathcal{N}=4$ sYM in the limit $N \to \infty$, while keeping the `t Hooft coupling $g^2 = g_{\rm YM}^2 N/(16 \pi^2)$ fixed \cite{tHooft:1973alw}.
In this limit, planar diagrams dominate, and diagrams of higher genus $G$ are suppressed by factors of $N^{-2 G}$. Unless otherwise stated, quantities in this and in the next section are assumed to be in this limit.
We can then write
\begin{equation}
\Gamma_{\rm cusp}(g) = \sum_{L \ge 1} g^{2 L} \Gamma_{\rm cusp}^{(L)} \,,
\end{equation}
for the perturbative expansion of the cusp anomalous dimension, and analogously for other quantities.

\begin{marginnote}[]
\entry{Planar limit}{Combined limit of the rank of the gauge group and of the coupling, so that planar Feynman diagrams dominate.}
\end{marginnote}

\subsection{Maximal und uniform transcendentality principle}
\label{section-maximal-weight}

Let us start by looking at perturbative results for the cusp anomalous dimension.
Its three-loop value in $\mathcal{N}=4$ sYM is \cite{Bern:2005iz}
\begin{eqnarray}\label{cuspN4sYM}
\Gamma_{\rm cusp} 
&=& \left(\frac{g^2_{\rm YM} N}{4 \pi^2} \right) - \frac{\pi^2}{12}    \left(\frac{g^2_{\rm YM} N}{4 \pi^2} \right)^2 + \frac{11\pi^4}{720}  \left(\frac{g^2_{\rm YM} N}{4 \pi^2} \right)^3 + \mathcal{O}(g^8_{\rm YM}) \,.
\end{eqnarray}
Looking at Equation \ref{cuspN4sYM} one notices an intriguing pattern. 
All transcendental constants appearing in this formula are instances of zeta values $\zeta_n = \sum_{k\ge 1} 1/k^n$, for example, $\zeta_2 =\pi^2/6$. Moreover, if one assigns transcendental weight, or `transcendentality' $n$ to $\zeta_n$, then the coefficients at $L$ loops have weight $2 L -2$.
\begin{marginnote}[]
\entry{Transcendental weight}{partly heuristic but useful property of transcendental constants and iterated integrals (see section \ref{section-Feynman-integrals}).
}
\end{marginnote}

It is instructive to compare Equation \ref{cuspN4sYM} to its corresponding result in QCD without quarks,
\begin{eqnarray}\label{cuspQCD}
\Gamma^{\rm{QCD}}_{\rm cusp} &=&  C_{R} \left(\frac{ g^2_{\rm YM} }{4 \pi^2} \right) +  C_{R} C_{A} \left(\frac{ g^2_{\rm YM} }{4 \pi^2} \right) ^2 \left( -\frac{\pi^2}{12} +\frac{67}{36} \right)  + \nonumber \\
&& +  C_{R} C_{A}^2 \left(\frac{ g^2_{\rm YM} }{4 \pi^2} \right) ^3  \left( \frac{11 \pi^4}{720} + \frac{11 \zeta_{3}}{24} -\frac{67 \pi^2}{216} + \frac{245}{96} \right)  
 + \mathcal{O}(g^8_{\rm YM})\,,
\end{eqnarray}
Here $C_{R}$ and $C_{A} = N$ are quadratic Casimir operators of $SU(N)$. $R$ refers to the representation of the fields under consideration. Setting $R=A$ for the adjoint representation, we see a remarkable feature of Eqs. \ref{cuspN4sYM} and \ref{cuspQCD}: the leading transcendental weight terms agree between ${\mathcal{N}}=4$ sYM and QCD \cite{Kotikov:2002ab,Kotikov:2004er}!

This agreement of the `most complicated terms'  was predicted based on an argument that the leading weight contribution to this quantity comes entirely from gluons \cite{Kotikov:2004er}. On the other hand, more general quantities, such as scattering amplitudes, may have maximal weight terms differing from those in $\mathcal{N}=4 $ sYM. 

Nevertheless, in retrospect the qualitative pattern that quantities in $\mathcal{N}=4$ sYM have maximal weight turned out to be very important.
The notion of weight generalizes to functions, as we will see in section 4.2 and 5. All evidence to date supports that $L$-loop scattering amplitudes in $\mathcal{N}=4$ sYM have uniform and maximal weight $2L$.

\subsection{Exact result for planar cusp anomalous dimension}

The cusp anomalous dimension is a prominent example of the application of integrability-based approaches in $\mathcal{N}=4$ sYM. 
As far as we are aware, there is no unambiguous or commonly agreed-upon definition of integrability in quantum field theory.
Usually it refers to a situation where (hidden) symmetries allow a problem to be solved exactly, to all orders. 
Here `exactly' could mean that the problem is recast in terms of a set of equations that in principle determine the answer (that may still involve complicated functions). 
One reason for thinking that $\mathcal{N}=4$ sYM theory may be integrable is the AdS/CFT correspondence, 
as signs of integrability are found in string theory \cite{Bena:2003wd,Dolan:2003uh}.

Following earlier work in QCD, reviewed in \cite{Belitsky:2004cz}, it was realized that anomalous dimensions of composite operators in the theory 
are equivalent to energies in certain integrable spin chain models \cite{Minahan:2002ve}. For example, at one loop the Heisenberg spin chain 
known from condensed matter physics makes an appearance. While the precise spin chain Hamiltonian was worked out explicitly only to low loop orders, it was the starting point for exploring integrability in planar $\mathcal{N}=4$ sYM. 
\begin{figure}[t]
\includegraphics[width=4in]{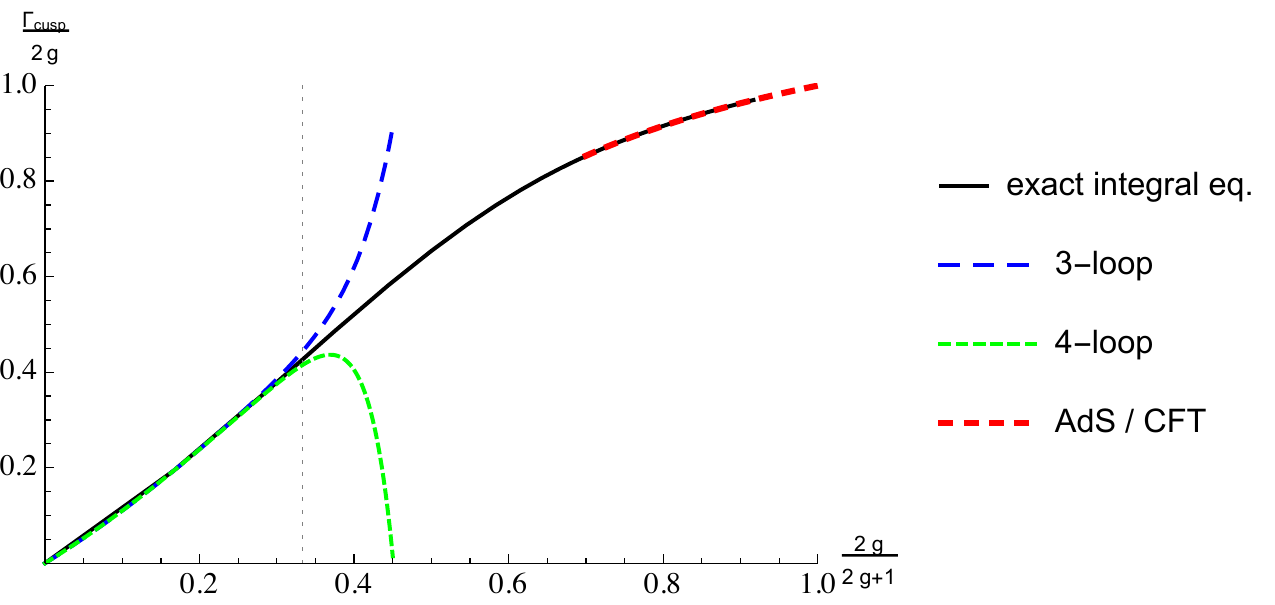}
\caption{Planar cusp anomalous dimension plotted on the whole range from weak to strong coupling, with $g^2 = g_{YM}^2 N/(16 \pi^2)$.
The numerical solution of an exact integral equation \cite{Beisert:2006ez} (solid line) agrees well with the three- and four-loop perturbative approximations, within the radius of convergence $g=1/4$ (vertical dashed line). It also agrees with the strong coupling expansion obtained from string theory. Figure adapted from \cite{Basso:2009gh}.
}
\label{fig1planarcusp}
\end{figure}
Assuming a Bethe ansatz inspired by integrability, the
cusp anomalous dimension is described by an integral equation valid to all orders in the coupling \cite{Beisert:2006ez}.
A truly impressive discovery!
The predictions of the latter agree with four-loop quantum field theory results \cite{Bern:2006ew}, as well as \cite{Benna:2006nd,Basso:2007wd}
with string theory results at strong coupling coupling \cite{Kruczenski:2002fb,Roiban:2007dq}, see Figure \ref{fig1planarcusp}.
Related to this there are first promising numerical results from a novel AdS/CFT lattice approach \cite{Bianchi:2016cyv}.

\subsection{First non-planar corrections}

Integrability results for the cusp anomalous dimension are currently limited to the planar limit.
Non-planar corrections to the cusp anomalous dimension appear for the first time at four loops 
and have recently been obtained analytically \cite{Henn:2019swt,Huber:2019fxe},  
\begin{eqnarray}\label{cuspN4sYM4loop}
\Gamma_{\rm cusp}
= \left( {\textrm{Eq.} \; 4} \right)
 - \left[ \frac{73 \pi^6}{20160}+\frac{\zeta_{3}^2}{8} + \frac{1}{N^2} \left( \frac{31 \pi^6}{5040} + \frac{9 \zeta_3^2}{4} \right) \right]  \left(\frac{g_{\rm YM}^2 N}{4 \pi^2} \right)^4 + \mathcal{O}(g_{\rm YM}^{10}) \,.
\end{eqnarray}
It is interesting to mention that taking into account known simpler contributions from fermions and scalars, the $\mathcal{N}=4$ sYM result 
provided the last missing ingredient to the full color dependence of the four-loop QCD cusp anomalous dimensions \cite{Henn:2019swt}, which was later reproduced by a direct QCD computation \cite{vonManteuffel:2020vjv}.

\section{AMPLITUDES IN $\mathcal{N}=4$ SUPER YANG-MILLS THEORY}
\label{section-amplitudes-adscft}

\subsection{Infrared divergences of amplitudes in the planar limit}

The study of the structure of infrared divergences in gauge theories has a long history.
A key concept is the idea of factorization.
Specifically, effects coming from soft and collinear emissions decouple, and hence can be written in a factorized form. 
The collinear terms are universal and independent of the scattering process. The soft terms can be represented by emission from Wilson lines tracing the initial and final particle directions.

As an example, consider some scattering process involving a virtual gluon exchanged between two on-shell legs.
In the region of integration where the momentum $k$ of the gluon becomes soft, i.e. $k \rightarrow 0$, the scattering amplitude factorizes as
\begin{equation}\label{eqfactorization}
  \begin{minipage}[h]{0.8\linewidth}
	\vspace{0pt}
	\includegraphics[width=.6\linewidth]{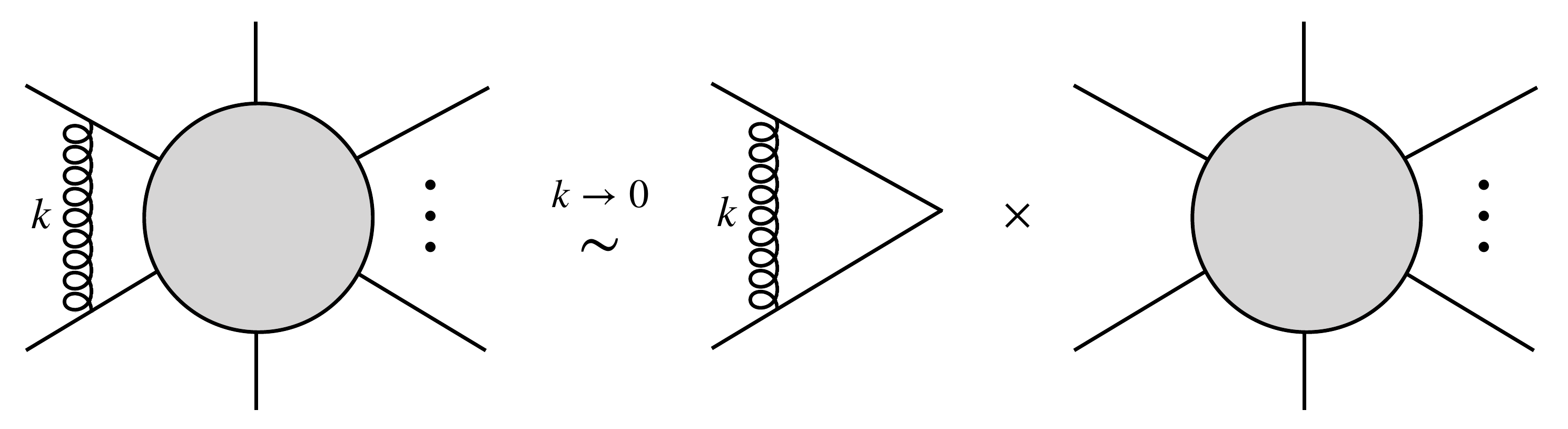}
		\vspace{-0.5cm}
    \end{minipage}
   \hspace{-0.5cm} \,.
\end{equation}
Here the blob denotes some hard scattering process (i.e. no further loop momenta are soft or collinear).
The physical meaning of Equation \ref{eqfactorization} is that soft gluons do not probe the hard scattering process.
The leading divergence of the vertex diagram on the right hand side of the equation is precisely the (one-loop value of) the cusp anomalous dimension.

Being an ultraviolet finite theory, ${\mathcal N}=4$ sYM is a particularly good testing ground for understanding infrared divergences. One encounters
them in their purest form, disentangled form ultraviolet divergences. 
In particular, in the planar limit, the divergences of an $n$-gluon amplitude $\mathcal{A}_{n}$, 
regulated dimensionally with $D=4-2 \epsilon$, take the form
\begin{equation}\label{Aplanardivergences}
\mathcal{A}_{n} = \mathcal{A}^{(0)}_{n} 
 \exp \left\{ 
 \sum_{L \ge 1} g^{2 L} \sum_{i=1}^{n}
 \left[ 
 - \frac{\Gamma_{\rm cusp}^{(L)}}{4 (L \epsilon)^2}
  - \frac{
  {\mathcal{G}}_{0}^{(L)}
  }{4 L \epsilon}
   \right]  \left( \frac{\mu^2}{s_{i\,i+1}} \right)^{L \epsilon} + F_{n}(g;s_{ij})  +
\mathcal{O}(\epsilon) 
 \right\}\,.
\end{equation}
Here $\mathcal{A}^{(0)}_{n}$ is the tree-level amplitude, ${\mathcal{G}}_{0}$ is the collinear anomalous dimension \cite{Dixon:2017nat}, and $F_{n}$ is the finite part.
$\mu^2$ is the dimensional regularization scale, and $s_{ij} = 2 p_{i} \cdot p_{j}$, where $p_{i}$ are the on-shell gluon momenta, and $s_{n\, n+1} = s_{1n}$.
We see that Equation \ref{Aplanardivergences} has a factorized structure \cite{Dixon:2008gr,Almelid:2015jia}. Its particularly simple form is due to the planar limit,
where soft/collinear exchanges can occur only between two particles at a given time. The latter lead to infrared divergences, which are represented by single and double poles in $\epsilon$ in Eq.~\ref{Aplanardivergences}.

\subsection{Amplitudes at strong coupling}
\label{sectionamplitudesstrongcoupling}

\begin{figure}
\centering
    \begin{minipage}{0.45\textwidth}
        \centering
        \includegraphics[width=0.65\textwidth]
                {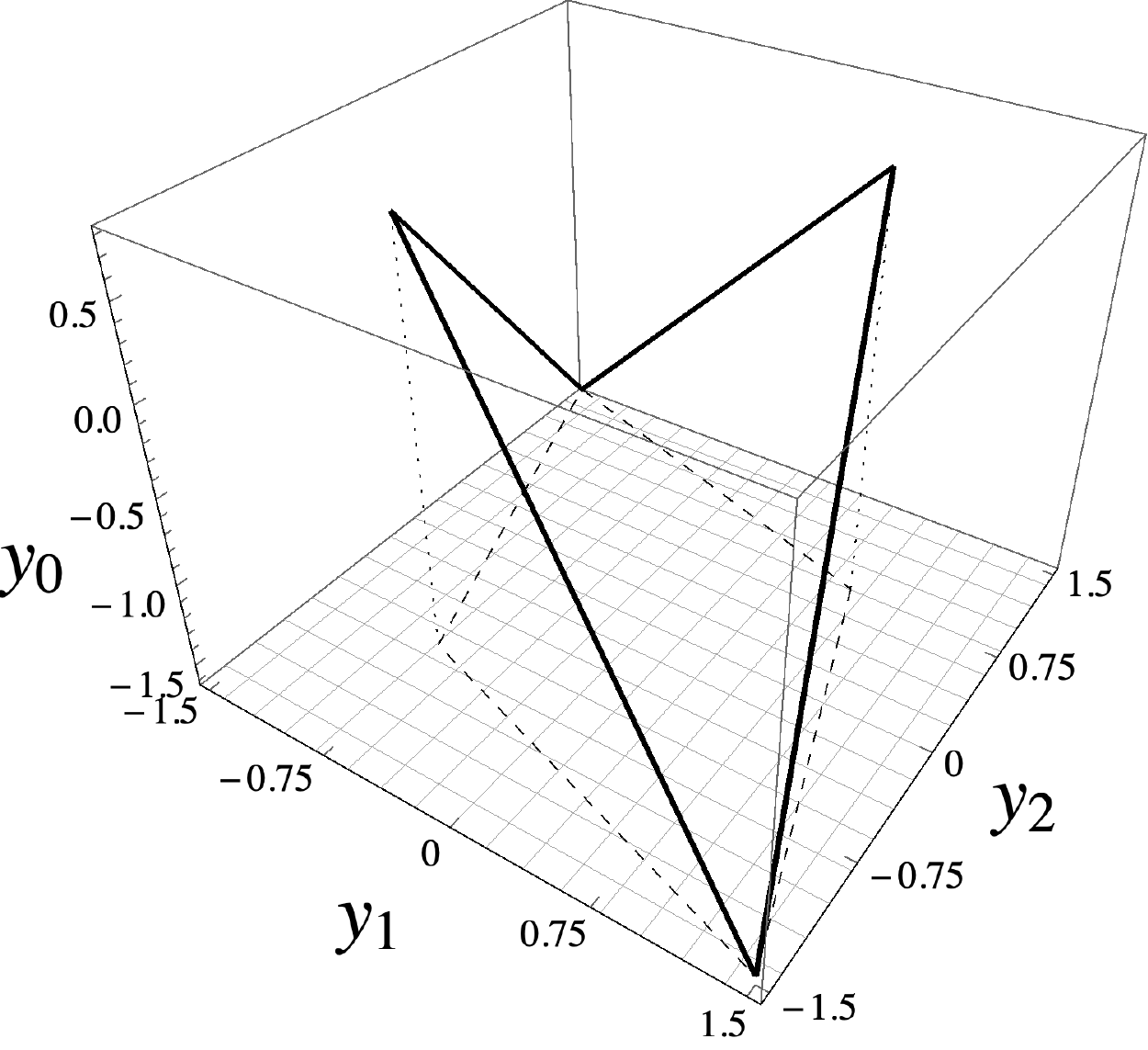} %
    \end{minipage}
    \hspace{0cm}
       \begin{minipage}{0.45\textwidth}
        \centering
        \includegraphics[width=0.8\textwidth]{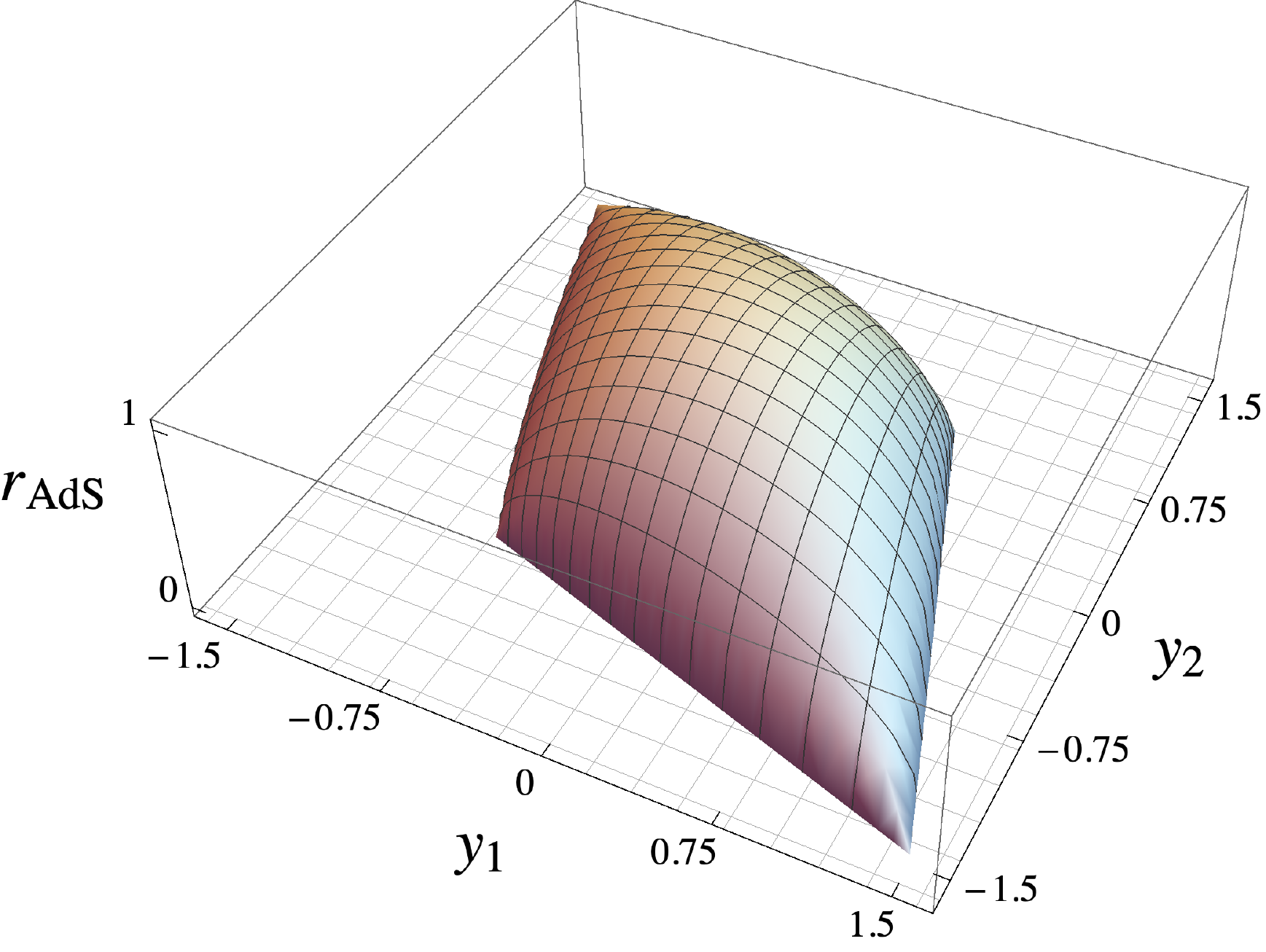} %
    \end{minipage}
\caption{
Left: Null polygon formed by the four gluon momenta, for $s/t=4$. 
Right: Minimal surface area solution of \cite{Alday:2007hr}, projected onto the $(y_1, y_2)$ plane.
The surface ends at $r_{\scriptsize\textrm{AdS}}=0$ on the null polygon. It extends into the radial direction, similar to a soap bubble. Both plots use Poincar\'e coordinates $(y^\mu ,r_{\scriptsize\textrm{AdS}})$.
}
\label{figamplitudesstrongcoupling}
\end{figure}

In the previous section we saw that the planar cusp anomalous dimension is known.
Surprisingly, the same is true for certain scattering amplitudes.
In \cite{Anastasiou:2003kj,Bern:2005iz}, based on three-loop calculations, an all-orders guess was put forward
for the finite part of the planar four-gluon amplitude in $\mathcal{N}=4$ sYM, 
\begin{equation}\label{ABDKBDS}
 F_{4}(g;t/s)   =  \frac{1}{4} \Gamma_{\rm cusp}(g)  \log^2(t/s) + C(g)  \,,
\end{equation}
with $s=s_{12}$ and $t=s_{23}$. 
This means that in addition to the exponentiation of the infrared divergences, the finite part also exponentiates in a very simple way.
Apart from the scheme-dependent constant $C(g)$, Equation \ref{ABDKBDS} predicts the full kinematic dependence of the amplitude, at any order of the coupling. 
In particular, Equation \ref{ABDKBDS} predicts the amplitudes at strong coupling.

The AdS/CFT duality relates observables in the gauge theory and in string theory. 
Until 2007, studies focused on correlation functions, anomalous dimensions and Wilson loops in the theory.
In a breakthrough paper \cite{Alday:2007hr}, it was shown that the computation of planar gluon scattering amplitudes at strong coupling is equivalent 
to a minimal surface area calculation in AdS${}_{5}$ space, with the surface ending on a polygon formed by the gluon momenta.
See Fig. \ref{figamplitudesstrongcoupling} for the four-particle case. 
The polygon is closed due to momentum conservation.
Amazingly, the regularized minimal surface area \cite{Alday:2007hr} agrees perfectly with Equations \ref{Aplanardivergences} and \ref{ABDKBDS}.

\begin{marginnote}[]
\entry{AdS${}_{5}$}{
Anti-de-Sitter space, whose boundary at $r_{\rm AdS}=0$ corresponds to four-dimensional Minkowski space-time.
}
\end{marginnote}

\subsection{Duality between scattering amplitudes and Wilson loops}
\label{sectionWLSAduality}

According to the AdS/CFT dictionary, a minimal surface area calculation corresponds to the vacuum expectation value of a Wilson loop.
Therefore the work of \cite{Alday:2007hr} suggested a duality between scattering amplitudes and Wilson loops at strong coupling.
Subsequently, the duality was found to hold also at weak coupling, first for the maximally-helicity-violating (MHV) \cite{Drummond:2007aua,Brandhuber:2007yx}, and later for general helicity configurations \cite{CaronHuot:2010ek}. 
The most non-trivial test is probably the agreement of the two-loop six-gluon MHV amplitude and the corresponding hexagonal Wilson loop \cite{Drummond:2008aq,Bern:2008ap}. 
Further evidence for the duality holding at all orders in the coupling was provided by a string theory argument in \cite{Berkovits:2008ic}. 
See \cite{Alday:2008yw} for a review.
\begin{marginnote}[]
\entry{MHV}{
The simplest non-trivial helicity configuration of gluon scattering amplitudes
in supersymmetric Yang-Mills theory.
}
\end{marginnote}

It is remarkable that scattering amplitudes can be computed from Wilson loops. 
The latter usually appear in special limits or in effective theories, such as 
heavy-quark effective theory \cite{Grozin:2004yc} and soft-collinear effective theory \cite{Bauer:2000yr}. 
What is truly remarkable here is that they describe not only the divergences, but also the finite part.

\begin{figure}[t]
\centering
    \begin{minipage}{0.45\textwidth}
        \centering
        \includegraphics[width=0.8\textwidth]{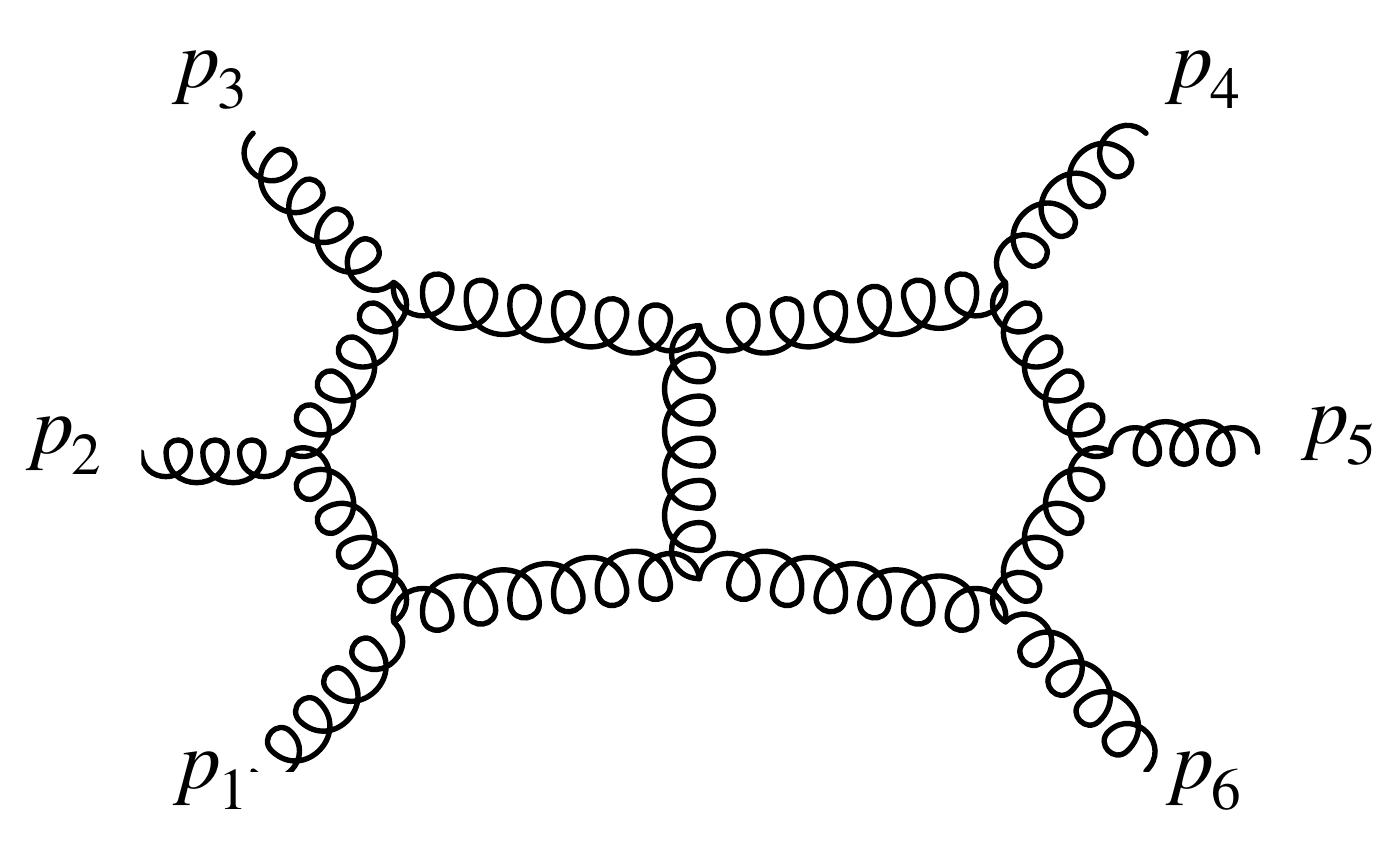} %
    \end{minipage}
    \hspace{0cm}
    \begin{minipage}{0.45\textwidth}
        \centering
        \includegraphics[width=0.4\textwidth]{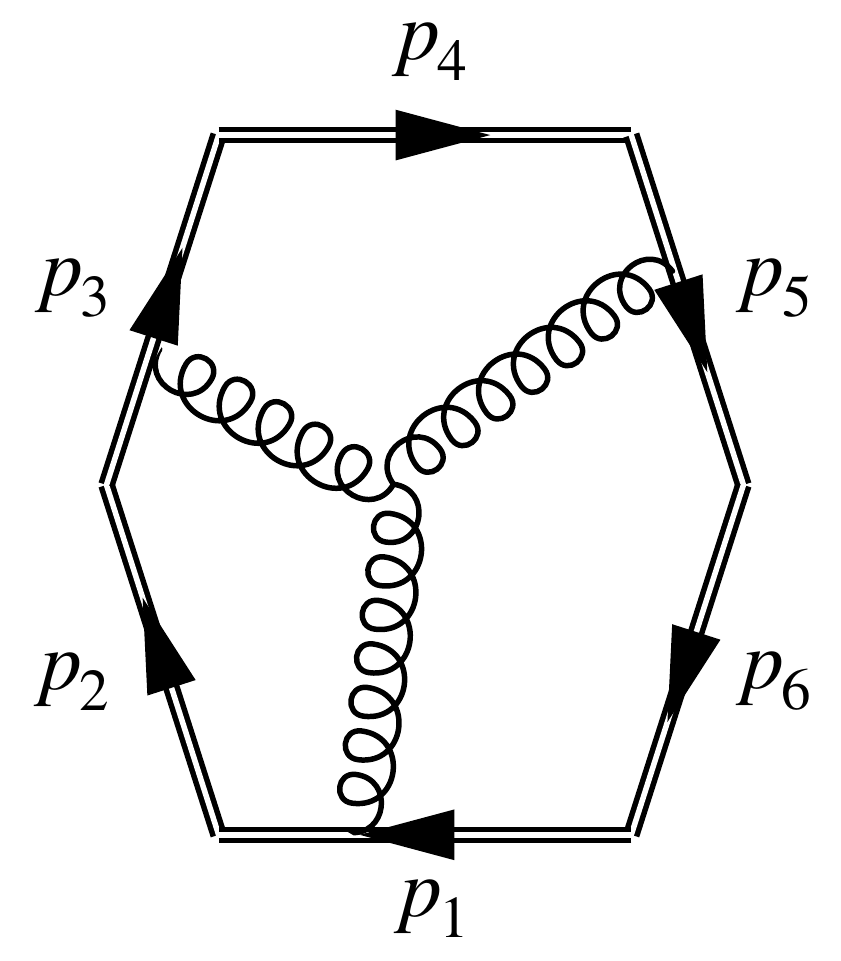}
    \end{minipage}
\caption{Sample Feynman diagrams contributing to the two-loop six-gluon amplitude (left), and to the two-loop hexagonal Wilson loop (right).}
\label{figdualitySA-WL}
\end{figure}

The Wilson loop picture offers several conceptual advantages.
Firstly, while the scattering amplitudes have infrared divergences, the Wilson loops have ultraviolet (cusp) divergences.
The structure of the latter is much easier to understand thanks to exponentiation properties of the eikonal lines \cite{Gatheral:1983cz}.
Secondly, the collinear limit of amplitudes corresponds to flattening one of the cusps in the Wilson loop picture. It turns out that not only the universal leading term, but also the near-collinear limit can be described using powerful operator product expansion techniques \cite{Alday:2010ku,Basso:2013vsa,Basso:2014hfa,Bonini:2015lfr}. 
Thirdly, at the practical level, Wilson loops are typically easier to evaluate than scattering amplitudes, so that the duality brings about technical simplification. This can be seen by the fact that one can evaluate numerically certain $n$-particle Wilson loops at two loops \cite{Anastasiou:2009kna} at any value of $n$. Much progress has occurred for the corresponding dual amplitudes, but currently their numerical evaluation in generic kinematics is limited to seven particles \cite{ArkaniHamed:2010gh,Henn:2018cdp,Bourjaily:2019vby,Dixon:2011nj}.

\subsection{Dual (super)conformal symmetry}
\label{sectionDCI}

Early hints at a hidden symmetry of scattering amplitudes were seen in special properties of the loop integrand \cite{Drummond:2006rz} of the four-gluon amplitude.
In the Wilson loop picture, the hidden symmetry is obvious: it is the coordinate-space conformal symmetry that transforms covariantly the polygon with light-like edges on which the Wilson loop is defined.
For the amplitudes, it becomes a dual conformal symmetry (acting in momentum space), in addition to their coordinate space conformal symmetry.

The symmetry can be used to make quantitative predictions. The cusp divergences of the Wilson loop break the symmetry slightly, 
but in a way controlled by all-order Ward identities \cite{Drummond:2007au}. 
The latter are very powerful: they fix the kinematic dependence for four and five gluons, in agreement with Equation \ref{ABDKBDS}. 
Moreover, the general $n$-gluon formula with $n>5$ is essentially given by the exponentiation of the one-loop result, with coefficient $\Gamma_{\rm cusp}$, plus a function of $3 n -15$ dual conformal invariants. This is to be compared with $3n-11$ variables without the symmetry. For six gluons, this means three instead of seven variables!

The new dual conformal symmetry is part of a dual superconformal algebra \cite{Drummond:2008vq,Berkovits:2008ic}.
When combined with the original superconformal symmetry of the Lagrangian, one obtains a Yangian algebra \cite{Drummond:2009fd}, which is a hallmark of integrability. Interestingly, some of the dual superconformal symmetry generators are also anomalous and lead to powerful relations between amplitudes involving different number of external legs and loop orders \cite{CaronHuot:2011kk}.
\begin{marginnote}[]
\entry{Yangian}{
Infinite-dimensional Hopf algebra that often appears in two-dimensional QFTs and in spin chain models.
}
\end{marginnote}

While most studies in $\mathcal{N}=4$ sYM focus on massless scattering amplitudes, there is a natural way of introducing masses within the AdS/CFT correspondence.
In this way, one can define infrared finite amplitudes that have an exact dual conformal symmetry \cite{Alday:2009zm}.

\subsection{Discussion}

Some readers might think that a mysterious hidden symmetry is quite far removed from reality
even more so in a conformal theory. 
However, upon closer inspection dual conformal symmetry is much more familiar than it may seem. 
The new generator it provides is equivalent \cite{Caron-Huot:2014gia} to the well-known Runge-Lenz vector. 
It is responsible for the regularity of planetary orbits in the Kepler problem in classical mechanics,
and it explains the simplicity spectrum of the hydrogen atom in quantum mechanics \cite{Pauli1926}. 
\begin{marginnote}[]
\entry{Runge-Lenz vector}{
Additional constant of motion describing the shape and orientation of an elliptical planetary orbit.
}
\end{marginnote}

Does the scattering amplitudes / Wilson loops duality and dual conformal symmetry extend to the non-planar sector \cite{Ben-Israel:2018ckc,Bern:2018oao}? Genuine non-planar corrections to amplitudes start at two loops, and although difficult to obtain, some results are available for further investigations \cite{Henn:2016jdu,Abreu:2018aqd,Chicherin:2018yne}. 

What can we learn from the fascinating results and dualities in $\mathcal{N}=4$ sYM for QCD? 
First, symmetries play an important role in the $\mathcal{N}=4$ sYM story. 
It would be interesting to disentangle the role of conformal symmetry from that of the extended supersymmetries. How powerful are the consequences of conformal symmetry alone, for example, for QCD at a conformal fixed point? 
Second, QCD and ${\mathcal N}=4$ sYM are more similar in singular limits. Does this indicate that these are universal properties of Yang-Mills gauge theories?
Third, having a Yang-Mills theory where results are known to high orders in perturbation theory, or even exactly, gives us a unique perspective on what may be possible in any gauge theory.
Indeed, many people would argue that the surprising simplicity of final results
obtained in $\mathcal{N}=4$ sYM suggests that there are better ways of thinking
about quantum field theory. These could well lead to new practical tools for QCD
calculations, and perhaps even to new perspectives on the foundations of quantum field theory.

\begin{textbox}[h]\section{THE MULTI-REGGE LIMIT OF QCD AND $\mathcal{N}=4$ SUPER YANG-MILLS}
QCD and supersymmetric Yang-Mills theories share the gluon sector, and in a qualitative sense this contains the most complicated diagrams.
In certain physical regimes, gluons also give quantitatively the main contributions.
This means that one can hope to find common properties in those limits.
This is the case for the multi-Regge limit, which plays an important role for multi-particle amplitudes in $\mathcal{N}=4$ sYM \cite{Bartels:2008ce}, see section \ref{section-bootstrapping-amplitudes}.
In QCD integrability was first seen in this limit, including a two-dimensional version of dual conformal symmetry \cite{Lipatov:1993yb,Faddeev:1994zg}. 
Moreover, while this is not the case for general kinematics, in this limit amplitudes in QCD are related to Wilson loops \cite{Korchemskaya:1996je}. 
\end{textbox}

\section{LESSONS FROM LOOP INTEGRANDS}
\label{section-integrands}

\subsection{On-shell methods}
\label{section-generalized-unitarity}

\begin{figure}[t]
\centering
    \begin{minipage}{0.75\textwidth}
        \centering
        \includegraphics[width=0.75\textwidth]{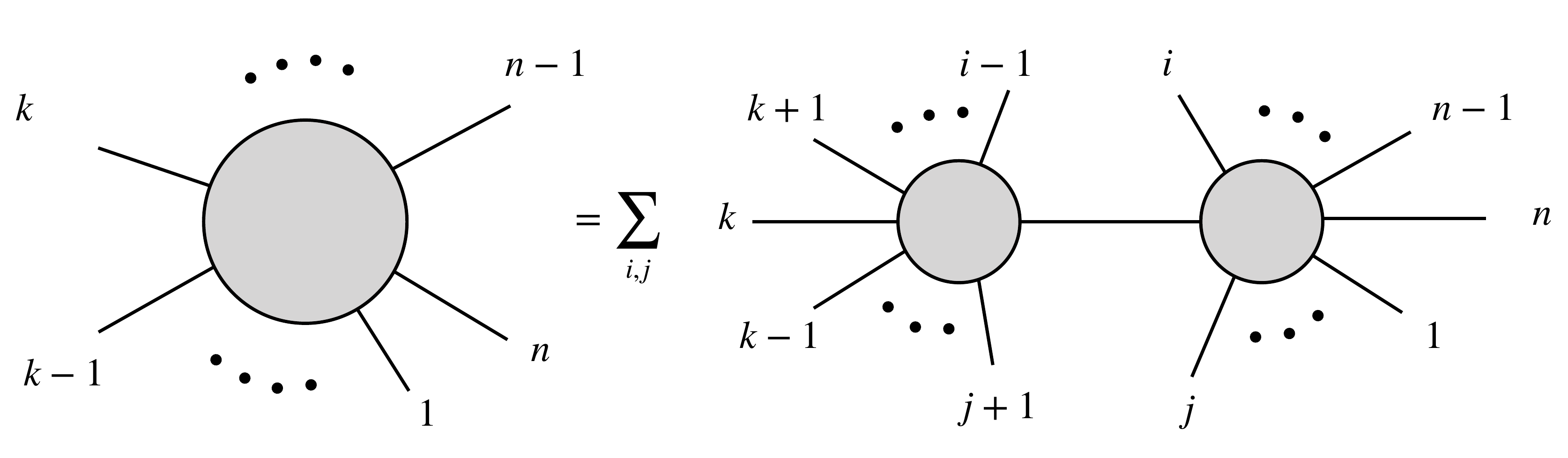} 
    \end{minipage}\hfill
\caption{On-shell recursion relations for tree-level amplitudes. Figure adapted from \cite{Britto:2005fq}.}
\label{fig-unitarity}
\end{figure}

An important conceptual principle in scattering amplitudes is to work as much as possible with gauge-invariant objects.
The reason is that individual Feynman diagrams are not gauge invariant, but scattering amplitudes -- sums of Feynman diagrams evaluated  on-shell -- are. 
For tree and one-loop diagrams, the off-shell expressions are often very complex, but they dramatically simplify when they are evaluated on-shell.

It turns out that it is possible to obtain information on complicated (higher-point and higher-loop) amplitudes from simpler ones via on-shell recursion relations.
This is perhaps clearest at the level of tree-level amplitudes, see Figure \ref{fig-unitarity}.
The authors of \cite{Britto:2005fq} proved recursion relations for tree-level amplitudes
from general factorization properties and complex analysis.
These principles imply that all amplitudes ultimately follow from elementary on-shell three-particle vertices. 
No off-shell quantities are required, and both the input and the output quantities are gauge invariant.
This not only facilitates the computation of the amplitudes, but also leads to representations that are naturally much more compact compared to what one would obtain from Feynman diagrams.
For example, in this way one obtains closed form formulas for all tree-level amplitudes that are manifestly invariant under dual conformal symmetry \cite{Brandhuber:2008pf,Drummond:2008cr}.
This also led to new analytic representations of tree-level amplitudes in massless QCD \cite{Dixon:2010ik}.

One important aspect in obtaining analytic insights is the use of appropriate variables.
In a landmark discovery, we learned that amplitudes have 
remarkably simple structures in twistor space \cite{Witten:2003nn}.
Moreover, momentum twistors \cite{Hodges:2009hk,Mason:2009qx} simultaneously solve the on-shell and momentum conservation constraints,
and are therefore unconstrained variables describing the kinematics. They have the additional benefit of transforming in a simple way under dual conformal transformation,
but can also be used very conveniently in models without that symmetry.

The tree-level on-shell recursion relations can be interpreted as a special case of a more general principle, namely perturbative unitarity.
By requiring unitarity of the $S$-matrix, and writing the latter as a sum of the identity (no interactions) and of an interacting part, one obtains relations between different loop orders.
Generalized unitarity, reviewed in \cite{Carrasco:2011hw}, takes this even further, and is
an efficient tool to find compact representations for loop integrands, bypassing the calculation of Feynman diagrams. 
\begin{marginnote}[]
\entry{Generalized unitarity}
{Method for obtaining the Feynman loop integrand from tree-level amplitudes.}
\end{marginnote}

The idea of using on-shell methods has been around for a long time, but only more recently came together into a general method for one-loop $n$-point scattering amplitudes, see e.g. \cite{Ossola:2006us,Berger:2008sj}. 
See \cite{Badger:2013gxa,Badger:2017jhb,Abreu:2018jgq,Dunbar:2019fcq} for very recent applications to two-loop QCD amplitudes.

Generalized unitarity methods to obtain loop integrands were refined to a point where it became possible to construct even non-planar loop integrands in various sYM theories at high perturbative orders. This means that in many cases, the bottleneck is now the evaluation of the loop integrations.
As a case in point, exploiting a relationship between color and kinematics \cite{Bern:2008qj}, and a connection between integrands in Yang-Mills and gravity \cite{Bern:2010ue}, Bern {\it et al} were able to obtain a five-loop integrand in $\mathcal{N}=8$ supergravity. To see how remarkable that is, one can consider that the number of terms contributing to it would have been bigger than the number of atoms in a desktop computer! Not only that, the authors were able to use this representation to test the ultraviolet behavior of supergravity amplitudes at that loop order \cite{Bern:2019prr}.

The way generalized unitarity usually is applied is that one makes an ansatz for the loop integrand and then applies various unitarity cuts to constrain the parameters in the ansatz until all of them are fixed. This approach has been very successful. But one can even go beyond this: in planar $\mathcal{N}=4$ sYM, it is possible to obtain the four-dimensional loop integrand recursively, generalizing the tree-level recursion relations \cite{ArkaniHamed:2010kv}. The same method is also expected to apply to other theories, but this requires resolving certain technical difficulties related to the definition of forward limits and the amputation of on-shell external legs for massless amplitudes. 
Remarkably, just as at tree-level, the on-shell recursion relations assemble the integrand from on-shell diagrams with only two elementary on-shell three-point vertices.
In other words, the information about the planar all-loop integrand in $\mathcal{N}=4$ sYM is essentially contained in on-shell diagrams.

A related way of thinking about on-shell diagrams is that they represent `maximal cuts' of loop integrands, where all loop integrations are replaced by residues. What one obtains in this way are so-called leading singularities of loop integrands. If one thinks of a generic term in an amplitude as $r \times f$, where $f$ is a multi-valued function, and $r$ is some rational or algebraic factor, then the idea is that taking residues of the integrand removes $f$ and leaves us with $r$.  While a quantitative connection is not known, it is expected that the leading singularities describe the factors $r$ that can appear in integrated amplitudes. 
This knowledge 
is especially important for bootstrap methods, see section 8.
\begin{marginnote}[]
\entry{Leading singularities}
{Multi-dimensional residues of loop integrands that localize all integrations.}
\end{marginnote}

Thanks to the impressive results above, it became possible to analyze the complete set of four-dimensional loop integrands in $\mathcal{N}=4$ sYM in detail.
It was found that on-shell diagrams can be conveniently described by a mathematical structure called positive Grassmannian \cite{Arkani-Hamed:2016byb}. 
This connection allows to classify and evaluate all on-shell diagrams, and to understand identities between them; moreover, their Yangian invariance is manifest.  
First studies for general theories include \cite{Franco:2015rma,Benincasa:2016awv}.

\subsection{Integrands with logarithmic singularities and uniform weight integrals}
\label{section-uniformweight-integrals}

It is useful to extend the notion of transcendental weight of section 2 to functions. 
This will be done more systematically in the next section, but
for now, let us assign weight $1$ to the logarithm, and weight $-1$ to $\epsilon$. 
All evidence to date suggests that $L$-loop Feynman integrals in four-dimensional QFT  evaluate to functions of weight $2 L$ at most.
Moreover, scattering amplitudes in ${\mathcal N}=4$ sYM appear to saturate this bound exactly, i.e. they are conjectured to be uniform weight $2 L$ functions, while amplitudes in QCD have mixed weights $0\le k\le 2 L$.
Why is this the case? 

\begin{figure}[t]
\centering
    \begin{minipage}{0.3\textwidth}
        \centering
        \includegraphics[width=0.7\textwidth]{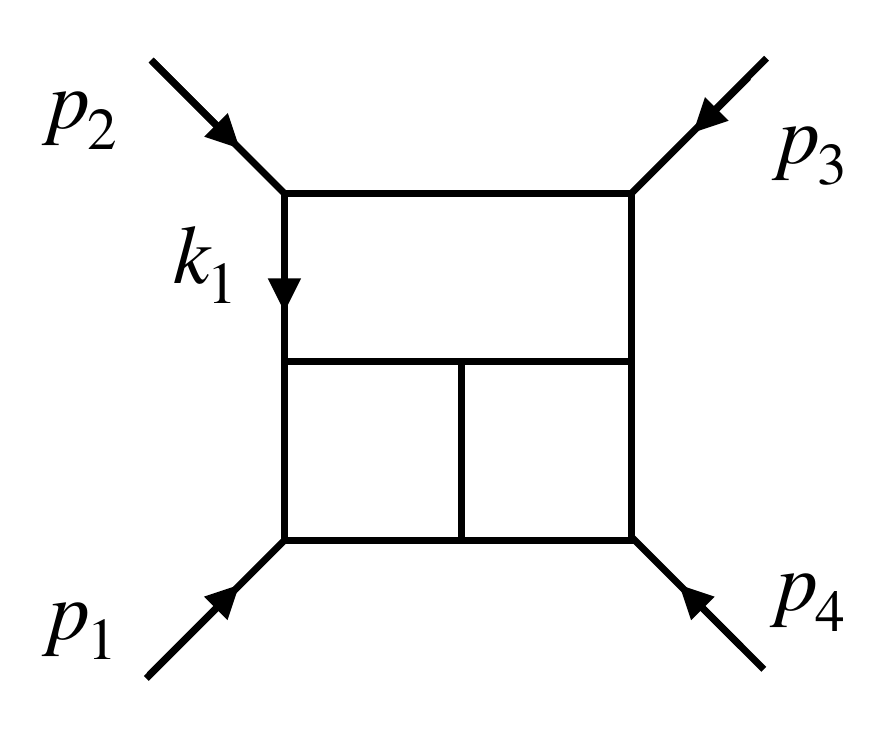} 
        (a)
    \end{minipage}
    \begin{minipage}{0.3\textwidth}
        \centering
        \includegraphics[width=0.7\textwidth]{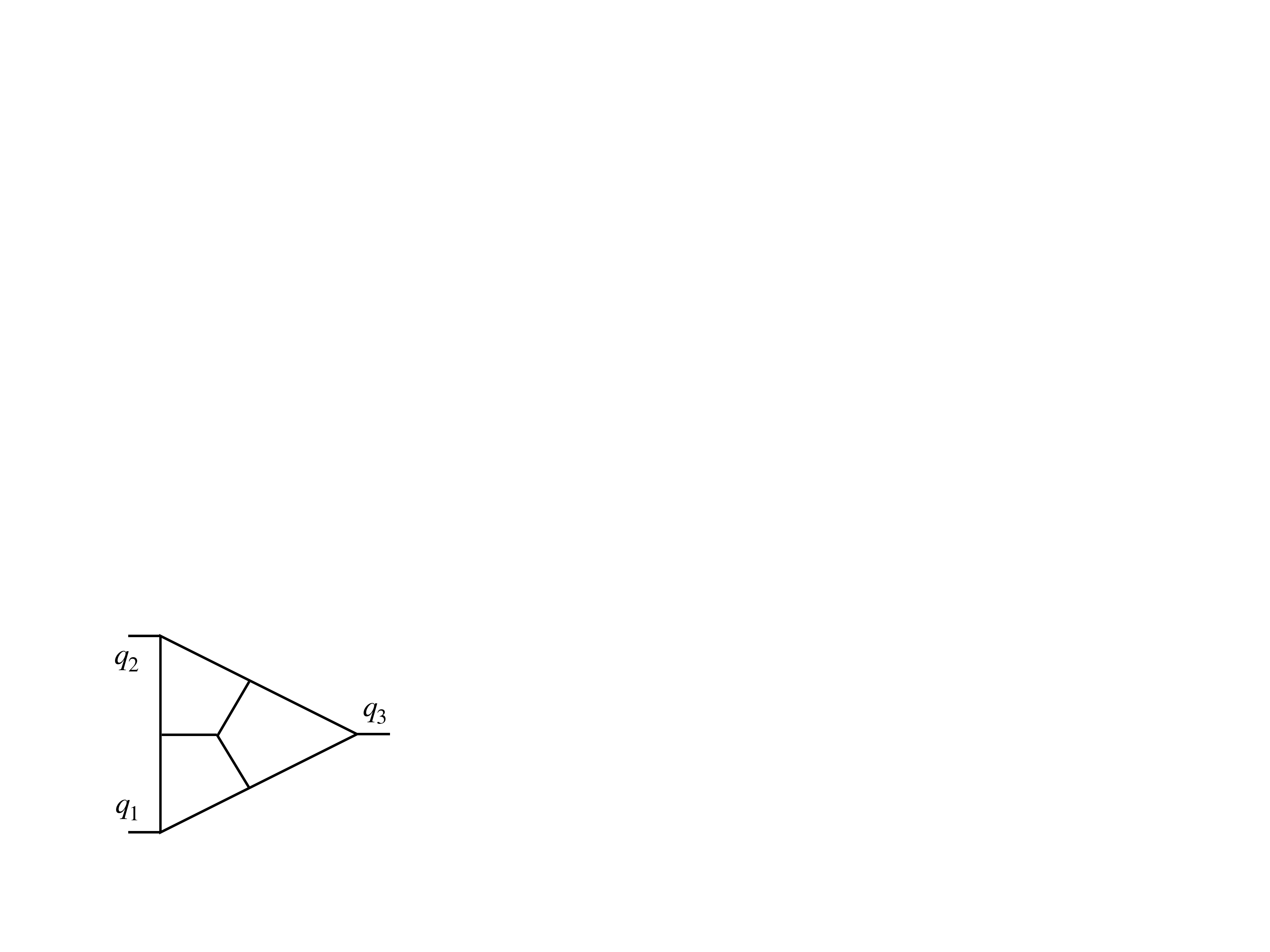} 
        \vspace{-0.15cm}
        (b)
    \end{minipage}
    \vspace{0.2cm}
       \caption{(a): Depending on its numerator, the Feynman diagram has different transcendental properties.
       (b): Off-shell dual conformal Feynman diagram with non-maximal weight. 
}
\label{figtenniscourtdiagrams}
\end{figure}

It turns out that the answer to the question lies in the loop {integrands}, i.e. the rational differential forms obtained from writing down Feynman diagrams. Apparently, in $\mathcal{N}=4$ sYM the Feynman diagrams conspire to give particularly `nice' integrands.
To appreciate the point, consider as an example the diagram in Fig.~\ref{figtenniscourtdiagrams}(a).
This is a particular three-loop Feynman integral obtained from $\phi^3$ vertices.
In $\mathcal{N}=4$ sYM, its contribution to the planar four-particle amplitude comes with a momentum-dependent numerator factor.
In the Lagrangian approach, the latter results from the sum of many Feynman diagrams.
What difference does it make?
To see this, let us inspect the first few terms in the Laurent expansion
 of the integrals in the two theories:
\begin{eqnarray}\label{eqYM}
        f_{{\rm sYM}} & =& (-s e^{ \gamma_{E}} )^{-3 \epsilon}\left[   - \frac{1}{\epsilon^6}\frac{16}{9} +  \frac{1}{\epsilon^5} \frac{13}{6} \log x + \mathcal{O}(\epsilon^{-4}) \right] \,,\\
        s^2 t^2 f_{\phi^3}  &=& (-s e^{ \gamma_{E}} )^{-3 \epsilon}\left[  \frac{1}{\epsilon^6}\frac{16}{9} + \frac{1}{\epsilon^5} \left( \frac{5}{6 x} + \frac{8}{9} -\frac{13}{6} \log x \right)   + \mathcal{O}(\epsilon^{-4})  \right] \,,
        \label{eqphi3}
\end{eqnarray}
where 
$x=t/s$,
While the $\mathcal{N}=4$ sYM result has uniform weight $2L=6$, the $\phi^3$ result is a more complicated mixture of weights, with additonal rational dependence on $x$.

This can be explained by inspecting the loop integrand. The basic idea is that differential forms of the type $d\tau/\tau$ lead to uniform weight functions, while double or higher poles do not. 
Note that `hidden' double poles can be revealed by taking (multiple) residues. 
The integrand of the $\phi^3$ graph of Figure~\ref{figtenniscourtdiagrams} contains double poles, while with the $\mathcal{N}=4$ sYM numerator it is a $4 L$-fold {\it dlog} differential form.
In general, computing leading singularities is an efficient way of testing the uniform weight property. This has been implemented algorithmically \cite{Henn:2020lye}, and 
applies to many situations, including non-planar integrals and integrals with massive propagators, which are essential for QCD phenomenology.

One might wonder whether dual conformal symmetry is equivalent to the uniform (and maximal) weight property, but in fact neither implication is true.
In the first direction, a counterexample is the three-loop integral shown in Figure~\ref{figtenniscourtdiagrams}(b). It is dual conformal, 
but it evaluates to ${20 \zeta_{5} }/({q_1^2 q_2^2 q_3^2 })$ and so does not have maximal weight six.
In the second direction, a counterexample is given by the one-loop triangle integral. 
In fact, it is often even possible to use a basis of uniform weight integrals, see section 5.

\begin{textbox}[h]\section{THE AMPLITUHEDRON: THE GEOMETRY OF FEYNMAN INTEGRANDS}
All evidence to date suggests that loop integrands in $\mathcal{N}=4$ sYM have only simple poles \cite{Arkani-Hamed:2014via}. 
In the case of planar MHV amplitudes, this statement follows from on-shell recursion relations. 
The authors of \cite{Arkani-Hamed:2013jha} take this further and propose a dual definition of the loop integrand of $\mathcal{N}=4$ sYM, the Amplituhedron: it is defined as the unique differential form that has logarithmic singularities (i.e. simple poles) only on the boundaries of a certain space related to the kinematics. 
This remarkable definition does not refer at all to the usual notion of local quantum fields.
On the one hand, unlike the Lagrangian formulation, where only conformal, but not dual conformal symmetry is manifest, 
in this formulation Yangian symmetry is built in. Moreover, it is free of gauge redundancy. 
On the other hand, concepts such as unitarity and locality appear as emergent properties.
\end{textbox}

\section{NOVEL METHODS FOR COMPUTING FEYNMAN INTEGRALS}
\label{section-Feynman-integrals}

An important problem in perturbative quantum field theory is the following:
given some rational loop integrand $\mathcal{I}$, consisting of products of propagators (and possibly some numerator factors coming from the Feynman rules), we would like to carry out the loop integrations,
\begin{equation}\label{eqFeynman1}
\mathcal{F}(p_{i};\epsilon) = \int \frac{d^{D}k_1 \ldots d^{D}k_{L}}{(i \pi^{D/2})^L} \mathcal{I}(k_i, p_j ) \,,
\end{equation}
where $k_i$ and $p_j$ are the loop and external momenta, respectively.
Typically, the integrals are regulated by the dimension $D=4-2\epsilon$, and we are interested in the answer in a Laurent series around $\epsilon=0$, which is truncated at some power of $\epsilon$.
It is useful to think of the answer as a sum of rational (or algebraic) factors $r$, and some special functions $g$, with constant coefficients $c$,
\begin{equation}\label{eqFeynman2}
\mathcal{F}(p_{i};\epsilon) = \sum_{k \le 2 L} \sum_{i,j} c_{ijk} \frac{1}{\epsilon^k}   r_{i} g_{j}  + \mathcal{O}(\epsilon)\,.
\end{equation}
The terms with inverse powers of $\epsilon$ represent ultraviolet or infrared singularities. 
In a practical cross-section calculation, the former terms are removed by renormalization, 
while the latter terms cancel against terms arising from integrating over the phase space of additional emitted particles according to the Bloch-Nordsieck theorem.
Here we used the heuristic information that the highest pole in $\epsilon$ is $2L$, and we truncated the expansion at the finite part, which is the physically relevant term.
Sometimes higher orders in $\epsilon$ are needed for consistency of the calculations, or are of genuine interest, for example in the method of $\epsilon$ expansion near a critical point.
\begin{marginnote}[]
\entry{Dimensional regularization}
{The space-time dimension is taken to be non-integer. Ultraviolet or infrared divergences appear as poles when the dimension approaches four.}
\end{marginnote}

What are the functions $g$ needed in quantum field theory, and specifically for scattering amplitudes?
In general, we expect multi-valued functions, since unitarity tells us that amplitudes have discontinuities at thresholds where particles are produced. 
One can show that, for any one-loop calculation up to $\mathcal{O}(\epsilon)$, the only special functions needed are the logarithm, and one further function, the dilogarithm. They are defined as
\begin{equation}\label{definitionlogandLi2}
\log(x) =\int_1^x \frac{dt}{t} \,,\quad\quad  {\rm Li}_{2}(x) = - \int_0^x \frac{dt}{t} \log(1-t) \,.
\end{equation}
The dilogarithm has the series representation ${\rm Li}_{2}(x) = \sum_{k \ge 1} x^k/k^2$, and we see that ${\rm Li}_{2}(1) = \zeta_2$.
This suggests a generalization of the notion of transcendental weight of section 2. 
In the context of iterated integrals with logarithmic integration kernels, i.e. of the form $d\log \alpha$, we will call weight the number of iterated integrations,
i.e. one for the logarithm, and two for the dilogarithm. 
This notion naturally generalizes to multiple iterated integrals that appear at higher loop orders.

\subsection{Symbol method}
\label{section-symbol}

Understanding the properties of the special functions appearing in QFT and in particular, their functional identities is very important for a number of reasons.
It is desirable to present results in a compact way, making manifest as many (physical) properties as possible.
This can be difficult if there are hidden relations between the functions. 
The same is true when analytically comparing two results from different calculations. 
Moreover, identities can be used for obtaining representations that are tailored to specific purposes: 
one representation might be well suited for a series expansion, but perhaps another one is better suited for numerical evaluation. 

The dilogarithm and its generalizations satisfy many identities, 
for example
\begin{equation}\label{exampleLi2identity}
{\rm Li}_{2}(x) + {\rm Li}_{2}(1-x)  + \log x \log (1-x) = \frac{\pi^2}{6} \,.
\end{equation}
A very useful mathematical tool is the `symbol' \cite{Goncharov:2010jf} of an iterated integral. 
In a nutshell, it retains the integration kernels in the definition of the iterated integrals, 
but discards the range of integration. 
This, together with elementary properties of the $d\log$ integration kernels, allows one to easily check equations of the type \ref{exampleLi2identity}, 
but also more complicated multi-variable and higher-weight versions of it. 

\begin{marginnote}[]
\entry{Alphabet}{The set of integration kernels, called letters, needed to define classes of iterated integrals. Key ingredient of the symbol.}
\end{marginnote}
Taken together, the set of all integration kernels appearing in a given quantity is called the alphabet. For example, in the case of ${\rm Li}_{2}(x)$, it is the set $\{d\log(x), d\log(1-x) \}$. In the interest of brevity one often writes only the arguments, i.e. $\{x,1-x\}$. 
Knowing the alphabet of a given problem is very important, as it encodes crucial analytic information about the functions. 
For example, zeros in the alphabet indicate possible, though perhaps spurious, singularities or branch cuts.

The symbol and the notion of alphabet have been used in numerous calculations in $\mathcal{N}=4$ sYM and QCD. 
If used appropriately from the outset they guarantee that results are written in a form where the structure of iterated integrals is manifest.

Alternatively, it allows us to find simplifications.
As an example, consider the vacuum polarization contribution to the two-loop muon anomalous magnetic moment \cite{Jegerlehner:2009ry},
\begin{eqnarray}\label{anomalousmoment1}
    \hspace{-1 cm}
  \begin{minipage}[h]{0.35\linewidth}
	\vspace{0pt}
	\includegraphics[width=.35\linewidth]{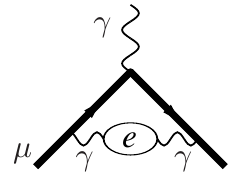}
    \end{minipage}
      \hspace{-1 cm}   &=& 
         \frac{x}{2}(1-5 x^2) 
         \underbrace{\left[ - \log x \log \left( \frac{x-1}{x+1} \right) + {\rm Li}_{2}(1/x) - {\rm Li}_{2}(-1/x)  \right]}_{f_1(x)} 
\nonumber \\
 & &      \hspace{-2.5 cm}   -\frac{25}{36} - \frac{\log x}{3} +x^2(4 +3 \log x) +  x^4 
 \underbrace{\left[ 2 \log^2 x- 2 \log x \log\left( x-\frac{1}{x} \right) + {\rm Li}_{2}(1/x^2) \right]}_{f_2(x)}    \,,
    \vspace{-0.3 cm}
\end{eqnarray}
where $x=m_{e}/m_{\mu}$.
Physically, this quantity should be single-valued for any value of $x>0$, but we see that individual terms have branch cuts starting at $x=1$. The symbol provides a quick way of verifying that those cuts cancel, and streamlines finding an equivalent expression with better analytic properties.
Using identities similar to Equation \ref{exampleLi2identity}, one finds
\begin{eqnarray}\label{anomalousmoment2}
f_1(x) =   \frac{\pi^2}{4}+ 2  {\rm Li}_{2}(1-x) - \frac{1}{2} {\rm Li}_{2}(1-x^2)  \,, \quad 
f_2(x) =  \frac{\pi^2}{6} + 2 \log^2 x + {\rm Li}_{2}(1-x^2)      \,.
\end{eqnarray}
In this form, the absence of branch cuts for $x>0$ is manifest.

The symbol is part of a coproduct structure. The latter helps in systematically keeping track of the integration constants, which are missed by the symbol.
We also wish to mention interesting recent work to find a coaction directly at the level of Feynman integrals \cite{Abreu:2017enx}.
\begin{marginnote}[]
\entry{Coproduct}{Map that decomposes iterated integrals into simpler building blocks.}
\end{marginnote}

\begin{textbox}[h]\section{SYMBOL ALPHABETS AND CLUSTER ALGEBRAS}
Perhaps the most striking use of the symbol was the simplification \cite{Goncharov:2010jf} of a previous expression \cite{DelDuca:2010zg} for the two-loop six-gluon MHV amplitude, reducing the size of the final answer from several pages to just a few lines!
For this, the observation that the full expression depends on fewer symbol letters than individual terms in the original expression, which suggested the use of a simpler function space. Remarkably, the alphabet identified by \cite{Goncharov:2010jf} turns out to be related to the intriguing mathematical structure of cluster algebras \cite{Golden:2013xva}. Cluster algebras are a relatively new topic in mathematics, with many fascinating links to topics in mathematics and physics. The connection to the alphabet, and the fact that cluster algebras also appear in the description of the planar loop integrand of $\mathcal{N}=4$ sYM \cite{Arkani-Hamed:2016byb} lends support to the conjecture that this alphabet is sufficient to describe six-gluon amplitudes at higher loop orders, see section \ref{section-bootstrapping-amplitudes}.
\end{textbox}

\subsection{Canonical differential equations method for computing Feynman integrals}
\label{section-canonicalDE}

In this section we will see that the notion of weight helps in computing Feynman integrals.
Uniform weight integrals satisfy simple differential equations. Let us illustrate this for the simple example of the dilogarithm. 
We define \begin{equation}
\vec{f}(x;\epsilon) =\left( \epsilon^2 {\rm Li}_{2}(1-x) , \epsilon \log x , 1 \right)^T \,,
\end{equation}
where we have introduced $\epsilon$ as a bookkeeping variable.
We assign weight $-1$ to $\epsilon$, so that all elements have homogeneous weight zero.
Then we have
\begin{equation}\label{toyexample}
\frac{\partial}{\partial x} 
 \vec{f}(x;\epsilon)
 = 
 \epsilon \left[  \frac{1}{x} \left( \begin{array}{rrr}
0 & 0 & 0 \\                                              
0 & 0 & 1 \\
0 & 0 & 0 \\                                              
\end{array}\right)
+
\frac{1}{1-x}
 \left( \begin{array}{rrr}
0 & 1 & 0 \\                                              
0 & 0 & 0 \\
0 & 0 & 0 \\                                              
\end{array}\right)\right]
 \vec{f}(x;\epsilon)\,.
\end{equation}
This, together with the boundary value $\vec{f}(1;\epsilon) = \{0,0,1\}$, uniquely determines $\vec{f}$.
This approach may appear very formal for such a simple example, however it is directly applicable to large classes of functions.
Indeed, the basic idea applies to any class of functions defined iteratively: $\epsilon$ plays the role of a bookkeeping variable
that keeps track of the number of integrations.

What does this have to do with Feynman integrals? We know that Feynman diagrams satisfy coupled differential equations \cite{Gehrmann:1999as}, but the form of the latter is complicated in general. So the idea is to switch to a basis of uniform weight integrals, if possible.
Then the above arguments apply, and we would expect those differential equations to simplify to give canonical differential equations \cite{Henn:2013pwa}.
In the case of multiple polylogarithms, those canonical differential equations read 
\begin{equation}\label{canonical-equation1}
d \vec{f}(\vec{x};\epsilon) = \epsilon \, \left[  \sum_i a_i\, d\log \alpha_i(\vec{x}) \right] \, \vec{f}(\vec{x};\epsilon) \,.
\end{equation}
Here $d=\sum_j dx_{j} \partial_{x_j}$, $\vec{f}$ has $M$ components, and the $a_{i}$ are constant $M \times M$ matrices.
Solving Equation \ref{canonical-equation1} in a series in $\epsilon$ defines a class of iterated integrals with integration kernels (letters) $\alpha_i$. 
The set of all letters is the alphabet. The solution at order $\epsilon^w$ has weight $w$. 
Our toy example of Equation \ref{toyexample} is a special one-variable case of Equation 
 \ref{canonical-equation1}, with alphabet $\{x , 1-x \}$ and $M=3$.
\begin{marginnote}[]
\entry{Multiple polylogarithms}
{Iterated integrals with logarithmic integration kernels. Also called hyperlogarithms.}
\end{marginnote}

Equation \ref{canonical-equation1} is valid for multiple polylogarithms, which cover many multi-loop cases. 
Sometimes even more complicated functions are required, such as multiple elliptic polylogarithms \cite{Duhr:2019wtr}. In this case, the  r.h.s. of Equation \ref{canonical-equation1} is still expected to be proportional to $\epsilon$, but more complicated integration kernels are needed. This is an active area of research.

In practice, for a given set of Feynman integrals, the method proceeds as follows. First, one finds uniform weight integrals, using the integrand analysis of section \ref{section-uniformweight-integrals}, but also via a number of semi-heuristic rules. Then, one choses a basis of uniform weight functions $\vec{f}$, and computes the differential equations, which should take the canonical form. In case one has only incomplete information about a uniform weight basis, a number of algorithmic tools can help find the remaining basis transformations \cite{Lee:2014ioa,Meyer:2016slj,Prausa:2017ltv,Gituliar:2017vzm,Dlapa:2020cwj}. 
For more information we refer interested readers to the lecture notes \cite{Henn:2014qga}.
This method has streamlined previously unattainable QCD computations that are crucial for next-to-next-to-leading order theoretical precision.

In summary, the symbol \cite{Goncharov:2010jf} and related coproduct techniques have become a standard tool in loop calculations, see e.g. \cite{Duhr:2012fh} and \cite{Duhr:2019wtr,Duhr:2019tlz} for useful computer codes. They are being used to analytically simplify expressions from previous calculations and to obtain representations that are well suited for fast numerical evaluation.

The canonical differential equation method \cite{Henn:2013pwa} has by now become an indispensable tool in QCD calculations relevant to collider physics. For example, it made possible new precise predictions for the production of two vector bosons \cite{Caola:2014lpa,Gehrmann:2014bfa,Gehrmann:2014fva}. 
The method led to a new degree of automation of calculations that brings within reach objectives that had previously appeared prohibitive, such as full next-to-next-to leading order corrections for processes involving many particles. 

\section{INFRARED FINITE OBSERVABLES}

\subsection{Infrared finite loop integrals and amplitudes}

Thanks to factorization (see section \ref{section-amplitudes-adscft}), the divergent terms in a given scattering process 
are in principle known beforehand. Indeed, the latter should cancel against real corrections when
going from scattering amplitudes to cross sections.
How this cancellation works technically depends on the details of the regularization procedure
and scheme, but the upshot is that the only new relevant information in a loop calculation
is the finite part, see e.g. \cite{Weinzierl:2011uz}.

It would therefore be desirable to be able to compute directly relevant finite observables. See \cite{Anastasiou:2018rib,Hannesdottir:2019rqq} for recent related discussions.
Being ultraviolet finite $\mathcal{N}=4$ sYM can help us understand better how to deal with the infrared divergences. Moreover, the exponentiation 
of divergences is conceptually simpler for Wilson loops. 

\begin{figure}[t]
\centering
    \begin{minipage}{0.8\textwidth}
        \centering
        \includegraphics[width=0.8\textwidth]{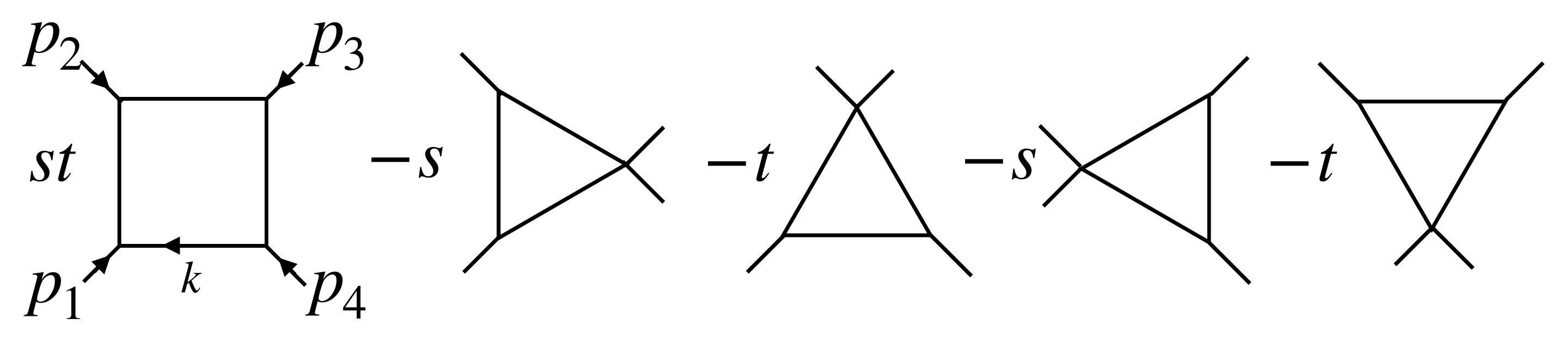} 
    \end{minipage}
       \caption{Finite linear combination of Feynman integrals.}
\label{figfiniteintegrals}
\end{figure}

Inspired by this, one can try to make the infrared properties as manifest as possible at the level of the loop integrand.
It turns out that generalized cuts are very useful in doing so: a subset of them identifies potentially singular regions, and by requiring them to be absent, one can impose the absence of (sub)divergences \cite{ArkaniHamed:2010gh}.
For example, consider a one-loop box integral with a numerator $N$,
\begin{equation}
\mathcal{F}_{4} = \int \frac{d^{D}k}{i \pi^{D/2}} \frac{ N(k ; p_{i}) }{k^2 (k+p_1)^2 (k+p_1+p_2)^2 (k-p_4)^2} \,,
\end{equation}
with
\begin{equation}
N(k; p_{i}) = s t \, - s (k-p_4)^2 - t (k+p_1+p_2)^2 - s (k+p_1)^2 - t k^2 \,.
\end{equation}
The first term in this sum is a scalar box integral, while the other terms are triangle integrals, see Figure~\ref{figfiniteintegrals}.
The numerator vanishes in all soft and collinear regions, e.g.
\begin{equation}\label{ircondition}
N(k = \alpha p_1 ; p_{i} ) = 0\,.
\end{equation}
Thanks to this property, the linear combination of box and triangle integrals is infrared finite. Indeed, one finds
\begin{equation}
\mathcal{F}_{4} = - \log^2( t/s ) - \pi^2\,.
\end{equation}
In general, imposing conditions such as Equation \ref{ircondition} allows one to control or predict the degree of (infrared) divergence of a given integral.
This was used to simplify QCD calculations of the cusp anomalous dimensions \cite{Henn:2019rmi}. See  \cite{vonManteuffel:2014qoa} for related work.
\begin{marginnote}[]
\entry{Collinear region}{scaling of loop momentum that upon integration may produce divergences.}
\end{marginnote}

Note that the knowledge of finiteness of the integral in Figure~\ref{figfiniteintegrals} can be used to make the infrared divergences of the box integral manifest.
That is, we can express them in terms of triangle integrals that depend only on one invariant, $s$ or $t$, in agreement with factorization.
The fact that this can be achieved at one loop is well known.
At higher loops, writing an amplitude explicitly as a sum of pieces with simple kinematic dependence and a finite remainder, is still an open problem, but the representations found in \cite{Bourjaily:2011hi,Bourjaily:2015jna} are progress in this direction. 

In addition, we would like to mention promising methods for evaluating finite Feynman integrals \cite{Caron-Huot:2014lda}, and
recent promising progress in computing Feynman integrals directly from their {\it dlog} form \cite{Lipstein:2013xra,Herrmann:2019upk}.
While we do not see a reason in principle why this method should not apply to dimensionally regulated integrals, it is certainly simpler for finite integrals. 

\subsection{Event shapes}

\begin{figure}[h]
\centering
    \begin{minipage}{0.45\textwidth}
        \centering
        \includegraphics[width=0.9\textwidth]{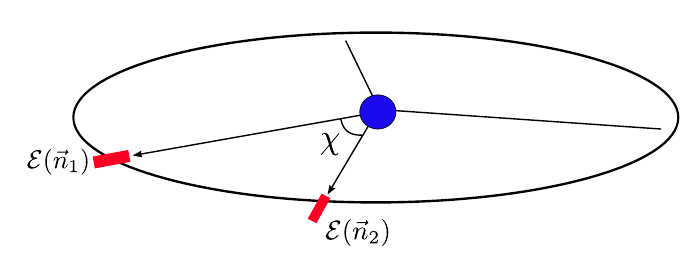} 
    \end{minipage}\hfill
\caption{Energy-energy correlation. Figure from \cite{Henn:2019gkr}.}
\label{figEEC}
\end{figure}

An important class of finite observables at hadron colliders are event shapes \cite{Basham:1978bw}.
An example is the correlation between energies deposited on calorimeters placed at a large distance from the collision, see Figure~\ref{figEEC}.
Event shapes have several nice properties, and may help to understand better the nature of jets in quantum field theory. 

Studying event shapes in $\mathcal{N}=4$ sYM is very interesting: from a practical point of view, the many special properties of the theory may allow computation at higher loop orders, thereby teaching us about structures to expect in QCD. 
But there are also conceptual advantages. Traditionally the energy-energy correlation is computed from scattering amplitudes which are divergent, and only at the level of the energy-weighted cross sections does one obtain a finite result.
Alternatively, event shapes can be defined starting from position-space correlation function \cite{Hofman:2008ar}.
This is particularly natural in a conformal field theory, such as $\mathcal{N}=4$ sYM.
Moreover, one may use operator product expansion techniques to infer properties of the event shapes \cite{Kologlu:2019mfz}.

The renewed interest in event shapes has already led to a number of remarkable results, both in $\mathcal{N}=4$ sYM and in QCD.
New QCD results include a novel analytic result of energy-energy correlators at next-to-leading order \cite{Dixon:2018qgp,Luo:2019nig},
and results on the collinear limit \cite{Korchemsky:2019nzm,Dixon:2019uzg,Kologlu:2019mfz}.
Furthermore, studies of multi-energy correlations have been initiated \cite{Chen:2019bpb,Chen:2020vvp}.
The method based on finite correlation functions was proven to be efficient in obtaining higher-loop corrections
in $\mathcal{N}=4$ sYM \cite{Henn:2019gkr}, and \cite{Chicherin:2020azt} provided a proof-of-concept application in QCD.

\section{BOOTSTRAPPING  
AMPLITUDES FROM THEIR ANALYTIC PROPERTIES}
\label{section-bootstrapping-amplitudes}

\subsection{The bootstrap philosophy: return of the analytic S-matrix program}
The analytic S-matrix bootstrap program has its origin in the 1960's \cite{Eden:1966dnq}. The idea is to constrain the S-matrix elements from general properties, such as unitarity, analyticity, and symmetries. 

In the area of scattering amplitudes, analytic properties play a very important role, as discussed in section 4. 
For example, one way of thinking about the  on-shell recursion relations for tree-level amplitudes 
is that they follow from the requirement that the amplitudes have the correct pole structure. 
At loop level, one encounters also branch cuts, which are related to unitarity of the S matrix. 

As early as 1994, in a groundbreaking paper, the authors of \cite{Bern:1994zx} constructed the full one-loop MHV $n$-gluon amplitudes in $\mathcal{N}=4$ sYM using a bootstrap philosophy. The critical pieces of information were understanding the allowed function space, and physical requirements such as the behavior in soft and collinear limits.

\subsection{Extension of the bootstrap ansatz to multi-loop amplitudes}

What are the difficulties in extending the bootstrap approach beyond the one-loop level? 
A priori, a given quantity in perturbation theory can be expanded according to Equation \ref{eqFeynman2}.
As a first simplification, if we focus on the finite part, we can remove the series in $\epsilon$ from that equation.
This leaves us with two key ingredients, the knowledge of the function space, and that of the coefficients.

In general, the function space of Feynman integrals is not fully understood beyond one loop. 
To illustrate, the state of the art is two-loop five-particle integrals for massless particles, or two-loop integrals in certain mass configurations.

In $\mathcal{N} = 4$ sYM, the situation is somewhat better. Thanks to dual conformal symmetry, the variable dependence
of planar amplitudes is significantly reduced. The first natural candidate for the bootstrap approach, beyond the known amplitudes,
is the planar MHV six-particle amplitude,
which is determined by a
function depending on three dual conformally invariant variables only, 
\begin{equation}
R\left( g^2; \frac{s_{12} s_{45}}{s_{123} s_{345}} ,
 \frac{s_{23} s_{56}}{s_{234} s_{123}} ,  \frac{s_{34} s_{61}}{s_{345} s_{234}} \right)\,.
\end{equation}

It turns out that there is a great deal of information about the key ingredients needed for the bootstrap technique.
A crucial breakthrough was made in \cite{Goncharov:2010jf} by recognizing that the function space needed to describe $R$ at two loops is characterized by a relatively simple nine-letter alphabet. Subsequently further evidence was collected that supports the idea that the same alphabet is sufficient at higher loop orders. 
In addition, it was shown in \cite{Arkani-Hamed:2016byb} that all (planar, four-dimensional) leading singularities for MHV amplitudes in $\mathcal{N}=4$ sYM are proportional to the tree-level amplitude, and it can be shown that the integrand has a {\it{dlog}} form. The last to two points suggest that the amplitude evaluates to a linear combination of weight $2 L$ functions, multiplied by the tree-level amplitude.

Taken together this leads to the following ansatz for the dual-conformally invariant part of the $L$-loop six-gluon amplitude \cite{Dixon:2011pw,Caron-Huot:2016owq,Caron-Huot:2019vjl},
\begin{equation}\label{eqbootstrapansatz}
R^{(L)} = \sum_{i} c_{i} h^{(2L)}_{i} \,,
\end{equation} 
where the tree-level amplitude has been stripped off, and hence, the $c_i$ are constants (i.e. kinematic-independent), and where $h^{(2L)}_i$ is the set of weight $2L$ hexagon functions, i.e. iterated integrals defined from the alphabet found by \cite{Goncharov:2010jf}. If we compare this with the general Equation \ref{eqFeynman2} makes it is clear how powerful the bootstrap input is. 

The next step is to use all available information to determine the $c_{i}$. 
The number of free parameters after application of each constraint is shown in Table \ref{tab1constraintsbootstrap}.
\begin{table}[h]
\tabcolsep7.5pt
\caption{Free parameters in the six-gluon amplitude, after imposing each constraint$^{\rm *}$.}
\label{tab1constraintsbootstrap}
\begin{center}
\begin{tabular}{| l |c|c|c|c|}
\hline
\cellcolor{gray!25}
Constraint \quad  \quad\quad\quad\quad\quad\quad\quad\quad\quad\quad\quad Loop order: &
 L=1 
 \cellcolor{gray!25} 
&
L=2 
\cellcolor{gray!25} 
&L=3 
\cellcolor{gray!25} 
&L=4
\cellcolor{gray!25} 
\\
\hline
Correct branch cut structure & 10 & 82  & 639 &5153 \\
\hline
Steinmann relations  \cite{Caron-Huot:2016owq}  &7 & 37 & 174 & 758\\
\hline
Cyclic and reflection symmetry & 3 & 11 & 44 & 174\\
\hline
Dual superconformal symmetry \cite{CaronHuot:2011kk} & 2 & 5 & 19 & 72\\
\hline
Universal factorization in collinear limits &0  &0 &1 &3\\
\hline
Constraints from multi-Regge kinematics \cite{Bartels:2008ce,Basso:2014pla} &0 &0 &0 &0\\
\hline
Near collinear limit from Wilson loop OPE \cite{Alday:2010ku} &0  &0 &0 &0\\
\hline
\end{tabular}
\end{center}
\begin{tabnote}
$^{\rm *}$Table adapted from Reference \cite{Caron-Huot:2016owq}. See also \cite{Caron-Huot:2019vjl}.
\end{tabnote}
\end{table}

\smallskip
It turns out that these constraints are so powerful that, depending on the loop order, applying some of them can fix the entire ansatz.
This means that further constraints give non-trivial consistency checks of the bootstrap assumptions, or provide predictions.

Let us discuss the first two constraints of Table~\ref{tab1constraintsbootstrap}. 
Planar massless amplitudes can have branch cuts (discontinuities) starting only at massless thresholds, such as e.g. $s_{12}=0$ or  $s_{123}=(p_1+p_2 + p_3)^2=0$.
This creates a condition for the hexagon functions entering the bootstrap ansatz. 
Similarly, the Steinmann relations \cite{Bartels:2008ce,Caron-Huot:2016owq,Caron-Huot:2019vjl} state that multiple discontinuities in crossed channels are forbidden, e.g.
\begin{equation}
   \hspace{-1cm}
     \begin{minipage}[h]{0.4\linewidth}
	\vspace{0pt}
       \includegraphics[width=0.4\textwidth]{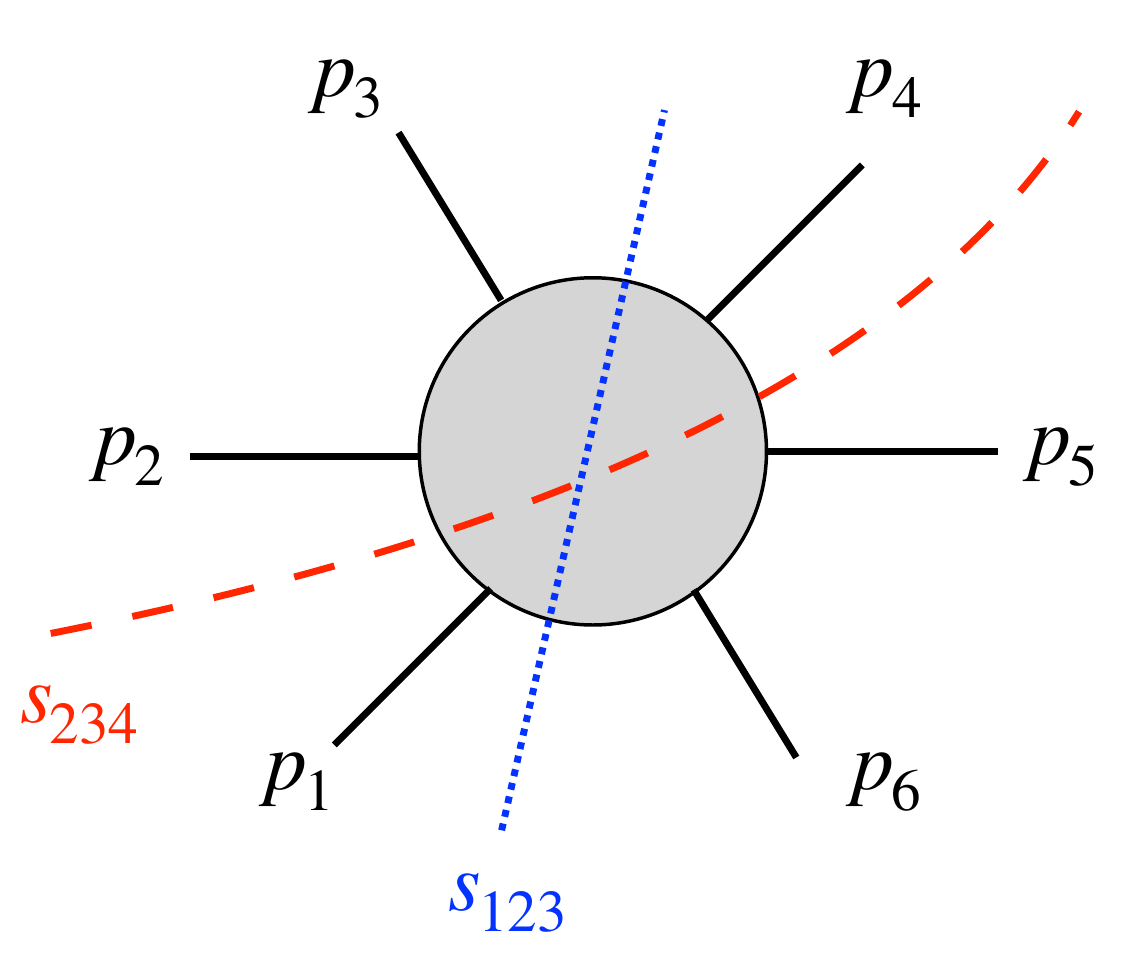}
    \end{minipage}
    \hspace{-1cm}
    \vspace{-0.4cm}
 = {\rm Disc}_{s_{123}} {\rm Disc}_{s_{234}} R = 0 \,.
\end{equation}
Interestingly, there is evidence that the Steinmann relations may be related to properties of cluster algebras \cite{Drummond:2017ssj}.

To date, this approach has been applied successively up to seven loops for six-gluon MHV amplitudes function \cite{Caron-Huot:2019vjl},
and impressive results are available for the six-gluon next-to-MHV helicity amplitude, and for the seven-gluon MHV amplitude \cite{Dixon:2016nkn,Drummond:2018caf}. The multi-variable functions yield much valuable information for analyzing quantum field theory amplitudes, for example by analyzing interesting kinematic limits.

In QCD, bootstrap ideas have already been used in a number applications.
This includes the rapidity anomalous dimensions \cite{Li:2016ctv}, and the soft anomalous dimension for multi-leg scattering  \cite{Almelid:2017qju}. 
Furthermore, the bootstrap can be a powerful tool for computing individual Feynman integrals \cite{Drummond:2013nda,Chicherin:2017dob}.

\section{DISCUSSION AND CONCLUSION}

In this article we have reviewed a selection of important advances for scattering amplitudes in $\mathcal{N}=4$ sYM, 
and developments in technology that were inspired by it. 
\begin{summary}[SUMMARY POINTS]
\begin{enumerate}
\item Planar four- and five-particle amplitudes are known exactly, in agreement with weak and strong coupling calculations.
\item Planar amplitudes in $\mathcal{N}=4$ sYM are described equivalently by polygonal Wilson loops. The conformal symmetry of the latter and properties in the near-collinear limit provide valuable information.
\item The bootstrap approach 
allowed to obtain planar six- and seven-gluon amplitudes in $\mathcal{N}=4$ sYM 
at impressively high loop orders.
\item Studies of the loop integrand in  $\mathcal{N}=4$ sYM, computed efficiently via on-shell methods, have uncovered unexpected analytic and geometric properties that hint at a dual formulation.
\item Observing that the singularity structure of loop integrands is related to the transcendental weight property has led to a novel method for computing Feynman integrals in any QFT.
\item Symbol methods for handling transcendental functions have become a standard tool for QCD calculations.
\end{enumerate}
\end{summary}

\subsection{How has $\mathcal{N}=4$ sYM knowledge fed into QCD calculations?}

Of course, the QCD and the $\mathcal{N}=4$ sYM communities are closely connected, and they profit from a constant exchange of ideas, for example through various conference series, such as Amplitudes, Loops and Legs, Loopfest, or Radcor, just to name a few. Given these fruitful interactions, it is of course impossible to neatly associate some ideas as definitely coming from one or the other communities. 
That said, many of the applications discussed in this review are arguably rooted more deeply in the $\mathcal{N}=4$ sYM side.

On the one hand, one of the most obvious uses is the similarity of calculations in perturbation theory at low perturbative orders, or in limits that expose universal behavior, such as soft, collinear or high-energy limits. A case in point is the three-loop soft anomalous dimension matrix.
On the other hand, in this review we focused more on $\mathcal{N}=4$ sYM as a fertile testing ground for new ideas and for discovering novel structures within QFT. 
We have seen several general purpose techniques, such as on-shell methods, techniques for Feynman integrands and integrals, that have had enormous impact on QCD calculations already. 
We think of the latter as the tip of an iceberg, as novel observations are continuously being made within this diverse and dynamic community of theorists.

\subsection{What more is possible?}

Let us comment on some future questions that may be interesting from the QCD viewpoint.
\begin{issues}[FUTURE ISSUES]
\begin{enumerate}
\item To what extent do the analytic and geometric structures observed in $\mathcal{N}=4$ sYM carry over to QCD integrands?
\item What is the precise form of canonical differential equations for Feynman integrals evaluating to functions beyond multiple polylogarithms?
\item To what extent do cluster algebras describe the QCD function space, and what can we learn from them? 
\item What extensions of the concept of four-dimensional leading singularities are necessary to describe the coefficients of transcendental functions in QCD?
\item $\mathcal{N}=4$ sYM provides a successful example of understanding the predictive power of (dual) superconformal symmetry in the context of infrared divergences. 
To what extent are massless QCD scattering amplitudes at the conformal fixed point determined by conformal symmetry? 
\end{enumerate}
\end{issues}
We hope that readers enjoyed learning about
some amazing adventures that are underway 
in the amplitudes community,
and that this review gives a fresh perspective on what is possible in a four-dimensional Yang-Mills theory, including some examples of surprising or unusual ways
of approaching problems in quantum field theory.

\section*{DISCLOSURE STATEMENT}
The author is not aware of any affiliations, memberships, funding, or financial holdings that
might be perceived as affecting the objectivity of this review. 

\section*{ACKNOWLEDGMENTS}
Sincere thanks to V.~Sotnikov, K.~Yan, S.~Zoia and especially to M.~Peskin for feedback on the manuscript, to G.~Korchemsky and V.~Forini for discussions, and to E.~Mendoza for linguistic advice.
This research received funding from the European Research Council  (ERC)  under  the  European  Union’s  Horizon  2020  research  and  innovation  programme, {\it{Novel structures in scattering amplitudes}} (grant agreement No 725110).

\bibliography{ar-bib}
\bibliographystyle{JHEP.bst}

\renewcommand\refname{RELATED PEDAGOGICAL REVIEWS}

\end{document}